\documentclass[twocolumn]{aastex61}

\newcommand{\ie}{i.e.}
\newcommand{\eg}{e.g.}
	
\newcommand{\kms}{km\,s$^{-1}$}
\newcommand{\teff}{$T_\mathrm{eff}$}
\newcommand{\logg}{$\log g$}
\newcommand{\feh}{[Fe/H]}

\newcommand{\vmic}{$\xi_\mathrm{micro}$}
\newcommand{\vmacro}{$v_\mathrm{macro}$}

\newcommand{\loggf}{log$(gf)$}
\newcommand{\mghi}{$^{24}$MgH}
\newcommand{\mghii}{$^{25}$MgH}
\newcommand{\mghiii}{$^{26}$MgH}
\newcommand{\mgi}{$^{24}$Mg}
\newcommand{\mgii}{$^{25}$Mg}
\newcommand{\mgiii}{$^{26}$Mg}
\newcommand{\cobold}{CO$^5$BOLD}
\newcommand{\lhdcode}{\texttt{LHD}}
\newcommand{\avg}{$\langle$3D$\rangle$}

\received{May 10, 2017}
\revised{June 11, 2017}
\accepted{June 12, 2017}
\submitjournal{ApJ}

\shorttitle{3D MgH line formation}
\shortauthors{Thygesen et al.}

\begin{document}

\title{An investigation of the formation and line properties of MgH in 3D hydrodynamical model stellar atmospheres}

\correspondingauthor{Anders O. Thygesen}
\email{thygesen@caltech.edu}

\author[0000-0002-4912-1183]{Anders O. Thygesen}
\affiliation{California Institute of Technology, 1200 E. California Blvd., Pasadena, CA 91125, USA}

\author{Evan N. Kirby}
\affiliation{California Institute of Technology, 1200 E. California Blvd., Pasadena, CA 91125, USA}

\author{Andrew J. Gallagher}
\affiliation{Max-Planck Institute  for Astronomy, K{\"o}nigstuhl 17, 69117 Heidelberg, Germany}
\affiliation{GEPI, Observatoire de Paris, PSL Research University, CNRS, Place Jules Janssen, 92190 Meudon, France}

\author{Hans-G. Ludwig}
\affiliation{Zentrum f\"{u}r Astronomie der Universit\"{a}t Heidelberg, Landessternwarte, K\"{o}nigstuhl 12, 69117 Heidelberg, Germany}

\author{Elisabetta Caffau}
\affiliation{GEPI, Observatoire de Paris, PSL Research University, CNRS, Place Jules Janssen, 92190 Meudon, France}

\author{Piercarlo Bonifacio}
\affiliation{GEPI, Observatoire de Paris, PSL Research University, CNRS, Place Jules Janssen, 92190 Meudon, France}

\author{Luca Sbordone}
\affiliation{European Southern Observatory, Alonso de Cordova 3107, Vitacura, Santiago, Chile}



\begin{abstract}
Studies of the isotopic composition of magnesium in cool stars have so far relied upon the use of one-dimensional (1D) model atmospheres. Since the isotopic ratios derived are based on asymmetries of optical MgH lines, it is important to test the impact from other effects affecting line asymmetries, like stellar convection. Here, we present a theoretical investigation of the effects of including self-consistent modeling of convection. Using spectral syntheses based on 3D hydrodynamical \cobold\ models of dwarfs (4000K\,$\lesssim$\,\teff\,$\lesssim$\,5160K, $4.0\leq$\,\logg\,$\leq4.5$, $-3.0\leq[\mathrm{Fe/H}]\leq-1.0$) and giants (\teff\,$\sim4000$K, \logg\,$=1.5$, $-3.0\leq[\mathrm{Fe/H}]\leq-1.0$), we perform a detailed analysis comparing 3D and 1D \textit{syntheses}. 

We describe the impact on the formation and behavior of MgH lines from using 3D models, and perform a qualitative assessment of the systematics introduced by the use of 1D syntheses. 

Using 3D model atmospheres significantly affect the strength of the MgH lines, especially in dwarfs, with 1D syntheses requiring an abundance correction of up to $+0.69$ dex largest for our 5000K models. The corrections are correlated with \teff\ and are also affected by the metallicity. The shape of the strong \mghi\ component in the 3D syntheses is poorly reproduced in 1D. This results in 1D syntheses underestimating \mgii\ by up to $\sim5$ percentage points and overestimating \mgi\ by a similar amount for dwarfs. This discrepancy increases with decreasing metallicity. \mgiii\ is recovered relatively well, with the largest difference being $\sim2$ percentage points. The use of 3D for giants has less impact, due to smaller differences in the atmospheric structure and a better reproduction of the line shape in 1D.

\end{abstract}

\keywords{Hydrodynamics -- Molecular processes -- Line: profiles - formation -- Stars: atmospheres -- Techniques: spectroscopic}



\section{Introduction} \label{sec:intro}

The field of elemental abundance studies of stars in the Milky Way (MW) has seen an enormous renaissance over the past decade with the advent of several large scale spectroscopic surveys like SEGUE \citep{yanny}, RAVE \citep{rave}, the Gaia-ESO survey \citep{gilmore}, GALAH \citep{2015MNRAS.449.2604D}, and APOGEE \citep{2015arXiv150905420M}, with several future surveys seeing first light within the next decade (\eg\ APOGEE-2; \citealt{apogee-2}, WEAVE; \citealt{weave}, PSF; \citealt{2014PASJ...66R...1T}, and 4MOST; \citealt{4most}), promising an unprecedented amount of spectroscopic data.

While the measurements of bulk abundances provides a lot of information about the chemical enrichment history of the MW, a deeper insight can be gained when one can also measure the isotopic composition of a given element. Most models of nucleosynthesis predict not only the bulk composition of various elements, but also the isotopic distribution. As such, measurements of the isotopic compositions provide more detailed information about the preceding nucleosynthesis. This is important if one is attempting to disentangle contributions from different sources of chemical enrichment, as different isotopes can be produced in different sites (see e.g. \citealt{2006A&A...456..313M} and \citealt{2010A&A...523A..24G} regarding Ba odd/even ratios).

In the case of low metallicity stars ([Fe/H$]\leq-1.0$ dex) isotopes of magnesium are of particular interest. The most abundant isotope, \mgi, is primarily produced in core-collapse supernovae, while the two heavier isotopes, \mgii\ and \mgiii, are predominantly produced in asymptotic giant branch (AGB) stars at low metallicities. This makes magnesium isotopes well suited for studying chemical enrichment timescales of old stellar populations like globular clusters (\eg\ \citealt{yong6752}, \citealt{melendezm71}, \citealt{dacosta}, and \citealt{thygesen}) as well as for stars belonging to the MW halo (\eg\ \citealt{mcwilliam}, \citealt{yongiso2003}, and \citealt{melendez07}). 

Since the isotopic splittings of the atomic lines of most elements are significantly smaller than the thermal line broadening in the stellar atmospheres, one has to rely on spectral line asymmetries in isotopic studies. Either in the atomic line themselves (\eg\ Li, Ba) or in molecular features (\eg\ CN, MgH, TiO). To accurately model the spectral line shapes it is paramount to include other potential causes for line asymmetries aside from the effect of different isotopes. 

Most isotopic studies have relied on the use of traditional 1D stellar atmospheres, which cannot model the convective motions of the gas in the stellar atmospheres. Convection in itself, can lead to line asymmetries as discussed in \eg\ \citet{1982ARA&A..20...61D}, \citet{1990A&A...228..184D}, \citet{2000A&A...359..729A}, \citet{2007A&A...473L..37C}, and \citet{2013MSAIS..24...78K}, which may mimic an isotopic signal. The use of full 3D model atmospheres resulted in a marked improvement for Ba odd/even ratios compared to 1D \citep{2015A&A...579A..94G}, as well as settling the debate about the existence of a $^6$Li plateau in metal-poor solar-type dwarf stars \citep{2006ApJ...644..229A,2010IAUS..268..191A,2012MSAIS..22..152S,2013A&A...554A..96L}.

In addition, the temperature fluctuations in the 3D hydrodynamical models, as well as differences in the overall temperature structure can have a dramatic impact on the line formation of diatomic molecules as showed by for instance \citet{2001A&A...372..601A}, \citet{2006ApJ...644L.121C}, \citet{2013MSAIS..24..138B}, \citet{gallagherCH}, and \citet{2015ApJ...806L..16B} for NH, CN, CH, OH and CO.

Recent improvements to the Linfor3D\footnote{\url{http://www.aip.de/Members/msteffen/linfor3D}} synthesis code \citep{gallagherLF3D} have now made it more practical to compute 3D syntheses over larger spectral regions. In this paper we present the first detailed investigation of the impact of using 3D hydrodynamical model atmospheres when modeling the optical features of MgH, typically used for the derivations of the distribution of Mg isotopes. 

\section{3D modeling of MgH features} \label{sec:modeling}

\subsection{The 3D model atmosphere grid}
To compute 3D syntheses of the optical MgH transitions we relied on a grid of models computed with the \cobold\ hydrodynamical atmosphere code \citep{co5bold}. The models were computed under the assumption of local thermodynamic equilibrium (LTE). The code computes a small region of the star (the so-called box-in-a-star mode), represented as a Cartesian box covering the upper part of the convective envelope and the stellar photosphere, with a typical optical depth of $-6<\log(\tau_{{\rm ross}})<7$. The physical size of the box changes with temperature and gravity, and is scaled from the standard model of the Sun ($5.6\times5.6\times2.3$ Mm), according to the pressure scale height at the surface, $\log(\tau_{{\rm ross}})=1$. 

The dwarf models considered here are a subset of the models computed by the ``Cosmological Impact of the First STars'' (CIFIST) collaboration \citep{2009MmSAI..80..711L}, while the giant models were computed for this project. Due to the substantial computational costs of constructing 3D hydrodynamical models, \cobold\ utilises precomputed opacity tables that have been grouped by wavelength \citep{1982A&A...107....1N}, with six opacity bins being the standard resolution. In the case of the giants, both standard and high-resolution models (12 opacity bins) were computed. The higher number of bins provides for a more accurate modeling of the outer layers of the stellar box, where the ranges in opacities can become very large. This may impact the spectral synthesis if the lines under investigation form predominantly in these layers. We compared the syntheses from six and 12 opacity bin models for the case of MgH, but did not find a noticeable difference. The 12 bin models were adopted for the remainder of the work presented in this paper.

We selected models in regions of the parameter space where one would expect to be able to detect Mg isotopes in MW halo stars. The full set of models are presented in Table~\ref{tab:hydrogrid}. Note that unlike in traditional 1D models, it is not possible to enforce a \teff\ for a 3D hydro model, as this depends on the exact realisation of the temperature fluctuations at a given instant in time. Hence the average model \teff\ will deviate somewhat from the formally desired \teff. 

In addition to the full 3D hydrodynamical models, two additional models are provided. For each 3D model instance, a so-called ``average 3D'' or \avg\ model is constructed by temporally and spatially averaging the temperature structure of the full 3D box over surfaces of equal Rosseland optical depth. The result is a 1D model with the same average thermal structure as the full 3D computational box. For details of the averaging procedure see \eg\ \citet{2012A&A...547A.118L}. Furthermore, a traditional 1D model was computed using the Lagrangian hydrodynamical (\lhdcode) code \citep{2007A&A...467L..11C}, which relies on the same input microphysics as \cobold. Specifically it uses the same equation-of-state and opacity binning scheme. Having these two sets of 1D models available allows us to directly investigate the impact of convection/temperature fluctuations (3D vs. \avg) and effects arising from changes to the overall temperature structure of the atmosphere (\avg\ vs. 1D \lhdcode). We will refer to the \lhdcode\ syntheses and models as ``1D'' for the remainder of the paper.

\begin{table*}
\caption{The \cobold\ model atmosphere grid used for our 3D syntheses. The models with the appended ``k12'' to the name indicates models computed with 12 opacity bins. For details on the derivation of the 1D model \vmic\ and \vmacro\ see Sect.~\ref{sec:analysis}. \label{tab:hydrogrid} }
\begin{tabular}{lccccc}
\hline \hline
       & & & & \multicolumn{2}{c}{1D only} \\
       \cline{5-6}
Model & \teff\ [K] &\logg\ [dex] &\feh\ [dex] & \vmic\ [\kms] & \vmacro\ [\kms] \\
\hline
d3t4000g150m10x140z150k12 & 4051 & 1.5 & $-1.0$ & 0.91 & 4.88 \\
d3t4000g150m20x140z150k12 & 4003 & 1.5 & $-2.0$ & 0.90 & 3.77 \\
d3t4000g150m30x140z150k12 & 4002 & 1.5 & $-3.0$ & 1.14 & 3.22 \\
d3t40g15mm10n01 & 4040 & 1.5 & $-1.0$ & 0.87 & 4.77 \\
d3t40g15mm20n01 & 4001 & 1.5 & $-2.0$ & 0.90 & 3.83 \\
d3t40g15mm30n01 & 3990 & 1.5 & $-3.0$ & 1.01 & 2.68 \\
\hline
d3t40g45mm10n01 & 4001 & 4.5 & $-1.0$  & 0.50 & 0.21 \\
d3t40g45mm20n01 & 4000 & 4.5 & $-2.0$  & 0.00 & 0.00 \\
d3t45g40mm10n01 & 4525 & 4.0 & $-1.0$  & 0.97 & 0.00 \\
d3t45g40mm20n01 & 4504 & 4.0 & $-2.0$  & 1.01 & 0.02 \\
d3t45g40mm30n02 & 4494 & 4.0 & $-3.0$  & 0.93 & 0.00 \\
d3t45g45mm10n01 & 4499 & 4.5 & $-1.0$  & 0.21 & 0.34 \\
d3t45g45mm20n01 & 4539 & 4.5 & $-2.0$  & 0.40 & 0.17 \\
d3t45g45mm30n01 & 4522 & 4.5 & $-3.0$  & 0.77 & 0.00 \\
d3t50g40mm10n01 & 4986 & 4.0 & $-1.0$  & 0.72 & 1.91 \\
d3t50g40mm20n01 & 4955 & 4.0 & $-2.0$  & 1.33 & 0.23 \\
d3t50g40mm30n02 & 5160 & 4.0 & $-3.0$  & 1.68 & 0.72 \\
d3t50g45mm10n03 & 5061 & 4.5 & $-1.0$  & 0.69 & 0.88 \\
d3t50g45mm20n03 & 5013 & 4.5 & $-2.0$  & 1.30 & 0.00 \\
d3t50g45mm30n03 & 4992 & 4.5 & $-3.0$  & 1.40 & 0.00 \\
\hline
\end{tabular}
\tablecomments{The model names for the dwarfs and the low-resolution giants follow the nomenclature used by the CIFIST collaboration. They indicate the ``desired'' model \teff\ (t), \logg\ (g), metallicity (m, the second ``m'' indicates that the metallicity is negative), and the model version (n). The names for the giant models with a high number of opacity bins follow largely the same naming scheme with model \teff\ (t), \logg\ (g), metallicity (m, would be ``p'' for positive metallicity), but are expanded to include number of grid cells in the horizontal directions (x,y), vertical direction (z), and number of opacity bins (k).}
\end{table*}

For the computation of the 3D syntheses a temporal sequence of computational boxes were selected from the full model sequence. We refer to these selections as snapshots. The snapshots were selected after the model had reached a statistically steady state. We used the 3D LTE synthesis code Linfor3D v.6.2.2 \citep{gallagherLF3D} for the computation of both the 3D, \avg\ and 1D syntheses. The syntheses for each 3D snapshot were subsequently averaged to yield the final 3D spectrum. Since only a small fraction of the stellar surface was simulated, we assumed that averaging over snapshots taken at different times was equivalent to averaging over the full stellar disk, as would be the case when dealing with spectra of real stars.

\begin{figure*}
\centering
\includegraphics[width=0.9\textwidth, trim = 3cm 4cm 3cm 3cm]{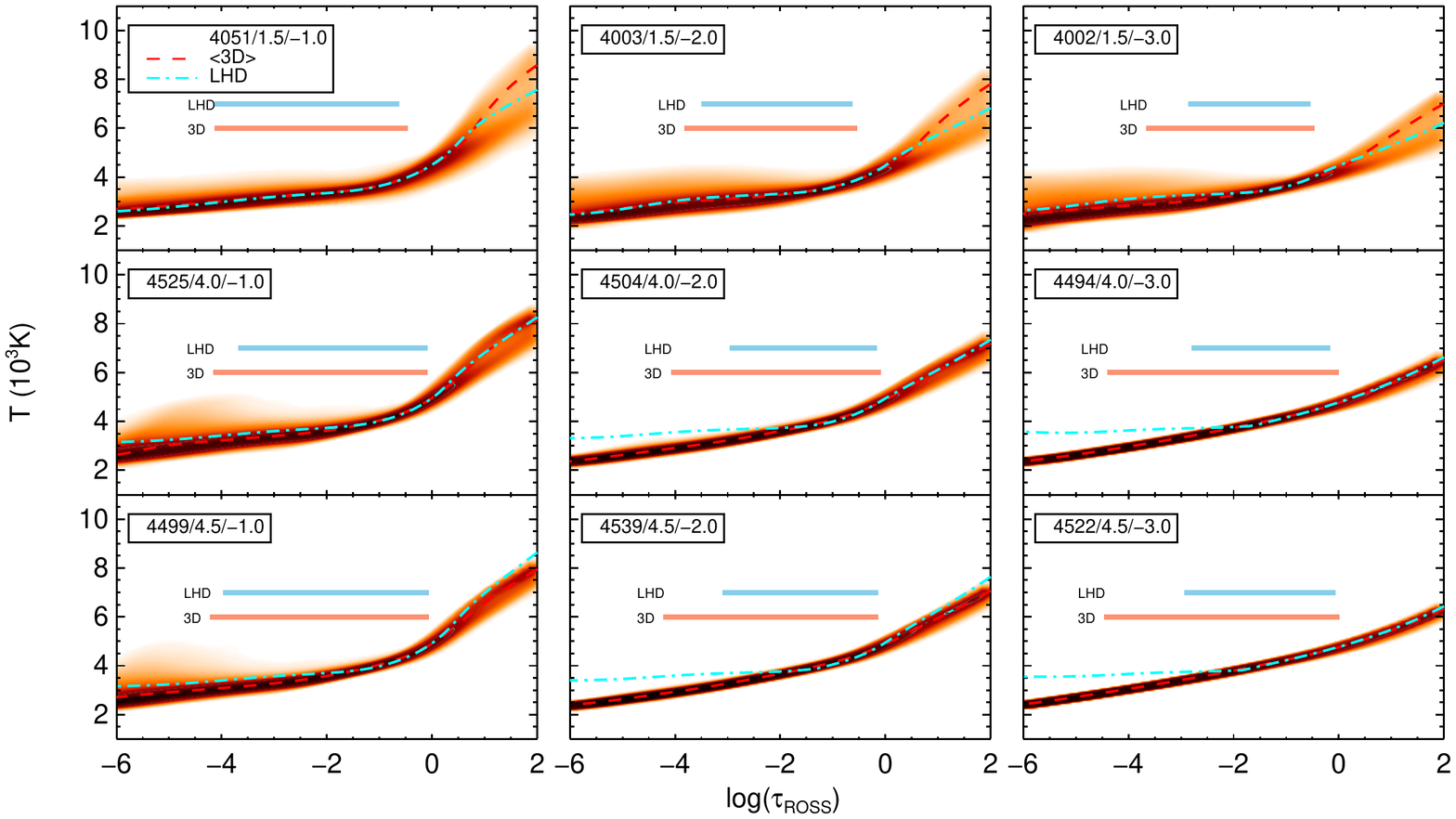}
\caption{Examples of the $T(\tau)$ structure of a subset of 3D \cobold\ models (density map) together with their \avg\ (red dashed line) and 1D (cyan dot-dash) counterparts. The darker areas indicate more frequently sampled temperature points. Shown for giants (top row), log$(g)=4.0$ dwarfs (middle), and log$(g)=4.5$ dwarfs (bottom). The model parameters are given in the captions. Note that each row defines a metallicity sequence. The formation regions of MgH in 3D (light red) and 1D (light blue) are indicated as horizontal bars. }%
\label{fig:t-structure}%
\end{figure*}

The temperature structure of 3D models can be markedly different from the traditional 1D models, especially at lower metallicities \citep{1990A&A...228..155N,2005ARA&A..43..481A,co5bold}. In Fig.~\ref{fig:t-structure} we present examples of the $T(\tau)$ structures of the full 3D computations of our 12 opacity bin giant models, as well as six dwarf models. The $T(\tau)$ structure of the remaining models are presented in Appendix~\ref{sec:ttau}, Figs.~\ref{fig:allTtau_4k} and \ref{fig:allTtau_5k}. In the case of the giants, relatively large temperature variations are observed in the line forming regions in 3D case, especially in the two low metallicity cases. This is in line with the more vigorous convection in these models, compared to a dwarf model. The \avg\ and 1D models, on the other hand, look very similar in terms of the overall structure in the line forming region, with the 1D giant model being slightly hotter in the $[\mathrm{Fe/H}]=-3.0$ case. 

The dwarf models, on the other hand, show very different model structures when the metallicity decreases. The 3D and \avg\ models have a notably cooler atmosphere than their 1D counterparts. This is the well-known extra cooling from the low amount of radiative line heating at the metallicities considered here, combined with the adiabatic expansion of the convective cells, which is not adequately modeled in 1D. This effect has been observed in most 3D models of low-metallicity dwarfs (see \eg\ \citealt{1999A&A...346L..17A} and \citealt{2008AIPC..990..268L}). Also the temperature fluctuations in the low-metallicity models are much less pronounced than in the high-metallicity case. This behavior is different than what has been observed in higher temperature dwarfs, where the temperature fluctuations tend to increase with decreasing metallicity \citep{gallagherCH,2013A&A...560A...8M}. However, there is evidence that this behavior reverses for the coolest dwarf models ($\lesssim5000$ K), with fluctuations increasing with increasing metallicity \citep{2013A&A...550A.103A,2013A&A...560A...8M}. 

\subsection{The input linelist}
\label{sec:linelist}

All of the MgH features contain blends from both atomic and molecular species. We constructed a line list for use in the synthesis using input from a number of sources. MgH transitions were taken from \citet{hinkle} and \citet{shayesteh}, with blending C$_2$, CN and CH lines from \citet{brooke,2014ApJS..210...23B} and \citet{masseron} respectively. In addition, we incorporated atomic blends from the Gaia-ESO line list \citep{ges}, supplied with atomic blends from VALD \citep{vald}. After having constructed an initial line list, we made a careful calibration by comparing spectral syntheses based on the standard \cobold\ model of the Sun, to a high-resolution spectrum of the Sun, taken from BASS2000\footnote{\url{http://bass2000.obspm.fr/solar_spect.php}} \citep{solar_atlas}. Based on the comparison with the Sun, we performed an astrophysical correction of the oscillator strength of the MgH lines at 5135.5\AA\ and 5140.2\AA\ to reproduce the solar spectrum. 

We used two different line lists, one for the MgH feature at 5135.1~\AA\ (188 transitions) and one for the two features at 5138.7~\AA\ and 5140.2~\AA, (409 transitions) since the two latter lines suffered from blends from the wings of two strong Fe-I lines at 5139.25~\AA\ and 5139.47~\AA, which needed accounting for in both cases. We attempted to minimize the number of lines in each line-list, but note that the majority of the transitions for each region were from either C$_2$ or CN, which may not leave a visible signature at the temperature of the Sun. Due to the significantly cooler temperatures of dwarfs targeted for MgH studies, molecular formation would be expected to be more efficient. For these reasons we decided to keep these transitions in our line list.

\section{Formation of MgH features} \label{sec:formation}

We computed a number of syntheses using a standard solar-scaled abundance pattern with $[\alpha/\mathrm{Fe}]=+0.4$ dex, and three isotopic compositions of magnesium (100:0:0, 94:3:3, and 88:6:6, expressed as percentage \mgi:\mgii:\mgiii). This is roughly the range that is expected for metal poor stars in the MW halo \citep{yongiso2003}.

\begin{figure*}
\centering
\includegraphics[scale=0.75, trim = 2cm 4cm 0cm 4cm]{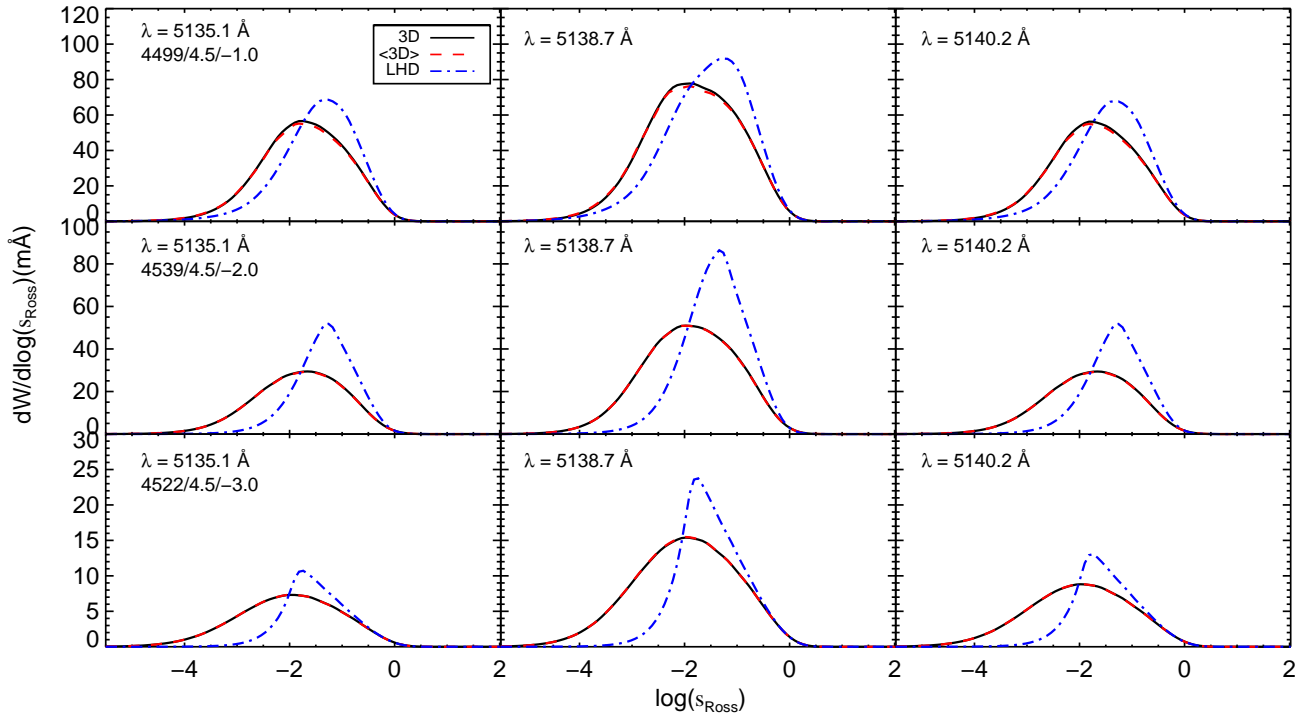}
\caption{The equivalent width contribution functions (MgH only) for the three MgH features under study. Shown for models d3t45g45mm10n01, d3t45g45mm20n01, and d3t45g45mm30n01 for both the 3D (black solid), \avg\ (red dash) and 1D (blue dot-dash) syntheses.}%
\label{fig:contf-mgh-dwarf}%
\end{figure*}

The differences in the temperature structure between 1D and 3D models (Fig.~\ref{fig:t-structure}) have a marked influence on the formation of the lines of MgH. In Fig.~\ref{fig:contf-mgh-dwarf}, we show the contribution functions of MgH only, for three representative dwarf models. The contribution functions have been computed with identical $[\mathrm{Mg/Fe}]=+0.4$, and an isotopic mixture of 88:6:6 in all cases shown here. They provide the fractional contribution to the line equivalent width (EW) as a function of optical depth, so that the integral gives the total line strength. The EW contribution functions were computed from the line depth contribution functions \citep{1986A&A...163..135M}. The formal definition of the EW contribution function, as implemented here, can be found in Eqn. (59) in the Linfor3D user manual\footnote{\url{http://www.aip.de/Members/msteffen/linfor3d}}. 

It is evident that when the metallicity decreases, the difference between using 3D models and 1D models becomes more substantial. In the 3D case, the region of line formation extends over a larger optical depth than is the case for the 1D models, with the contribution functions becoming shallower and broader. Furthermore, in 3D the line formation peaks in shallower layers than when using traditional atmospheres. These layers will have more efficient molecular formation as they are generally cooler in 3D, resulting in stronger features, becoming more pronounced as the metallicity decreases.

The effects of temperature fluctuations, on the other hand, seem to have negligible impact on the MgH line formation in the cases shown here. This is clear from comparing the contribution functions for the 3D and \avg\ cases, which are near-identical. Only in the $-1.0$ metallicity case is there a marginal difference between the full 3D and \avg\ contribution functions. This is also the model where the $T(\tau)$ structure (Fig.~\ref{fig:t-structure}, bottom row) exhibits the largest temperature fluctuations. It follows that the dominant reason for differences in line forming regions arises from differences to the overall structure of the atmospheres, as is evident by comparing 1D and \avg, rather than from $T$ fluctuations. 

\begin{figure*}
\centering
\includegraphics[scale=0.75, trim = 2cm 4cm 0cm 4cm]{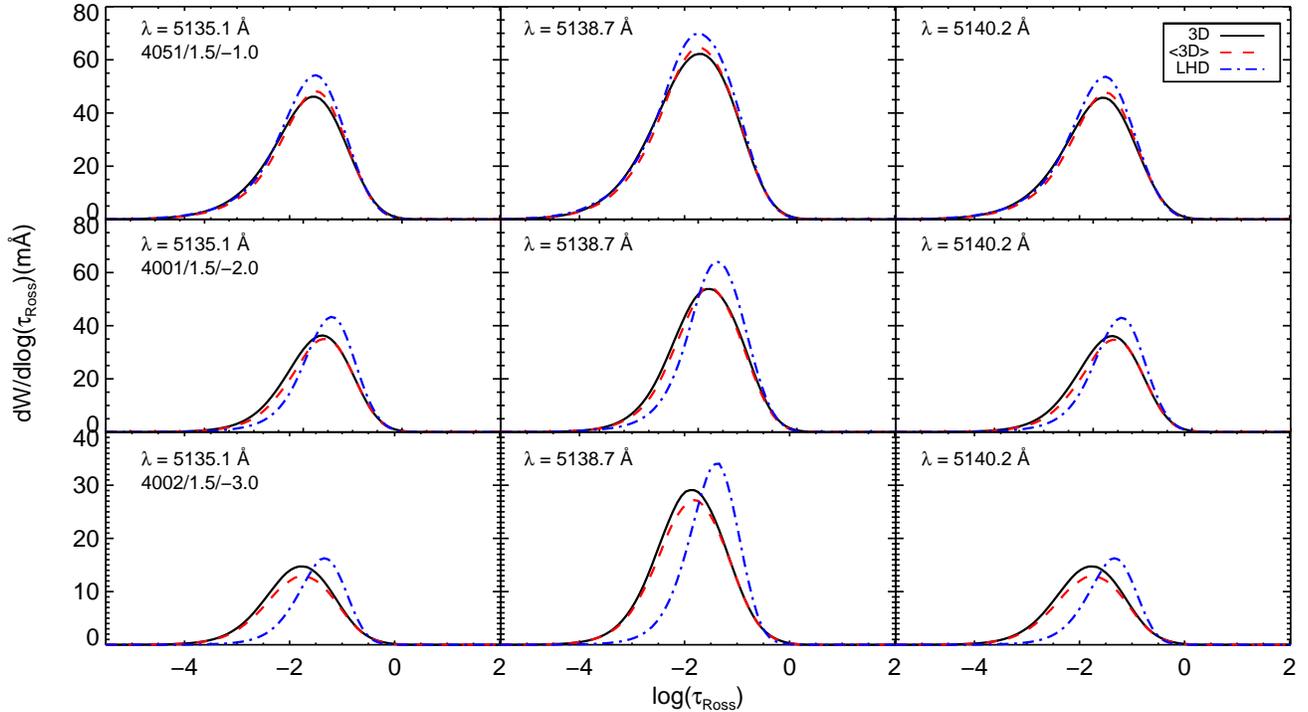}
\caption{As Fig.~\ref{fig:contf-mgh-dwarf}, but for the giant case.}%
\label{fig:contf-mgh-giant}%
\end{figure*}

In the case of giant stars (Fig.~\ref{fig:contf-mgh-giant}), some obvious differences are observed compared to the dwarf case. While the shift in the main formation depth was evident at all metallicities in the dwarfs, this is negligible for the giants at \feh~$=-1.0$, and is only becoming comparable to the dwarf case at the lowest metallicity. For the giants, the contribution functions are also notably shallower at high metallicity, indicating that the 3D MgH lines will tend to be weaker than their 1D counterpart. The influence of temperature fluctuations is more pronounced for giants than for the dwarfs investigated above. This is particularly true at the lowest metallicity, as suggested by the behavior of the $T$ fluctuations in the models. But as in the dwarf case, overall changes to the $T$ structure appear to be the dominant effect, becoming more important as the metallicity decreases. Although the $T(\tau)$ structure of the giants are very similar even at a metallicity of $-3.0$, it can be seen from Fig.~\ref{fig:t-structure} that there still is a difference of a few hundred K between the 3D and 1D models at \feh~$=-3.0$, which explains part of the differences seen in the contribution functions.

\begin{figure*}
\centering
\includegraphics[width=0.9\textwidth, trim = 2cm 3.5cm 0cm 3.5cm]{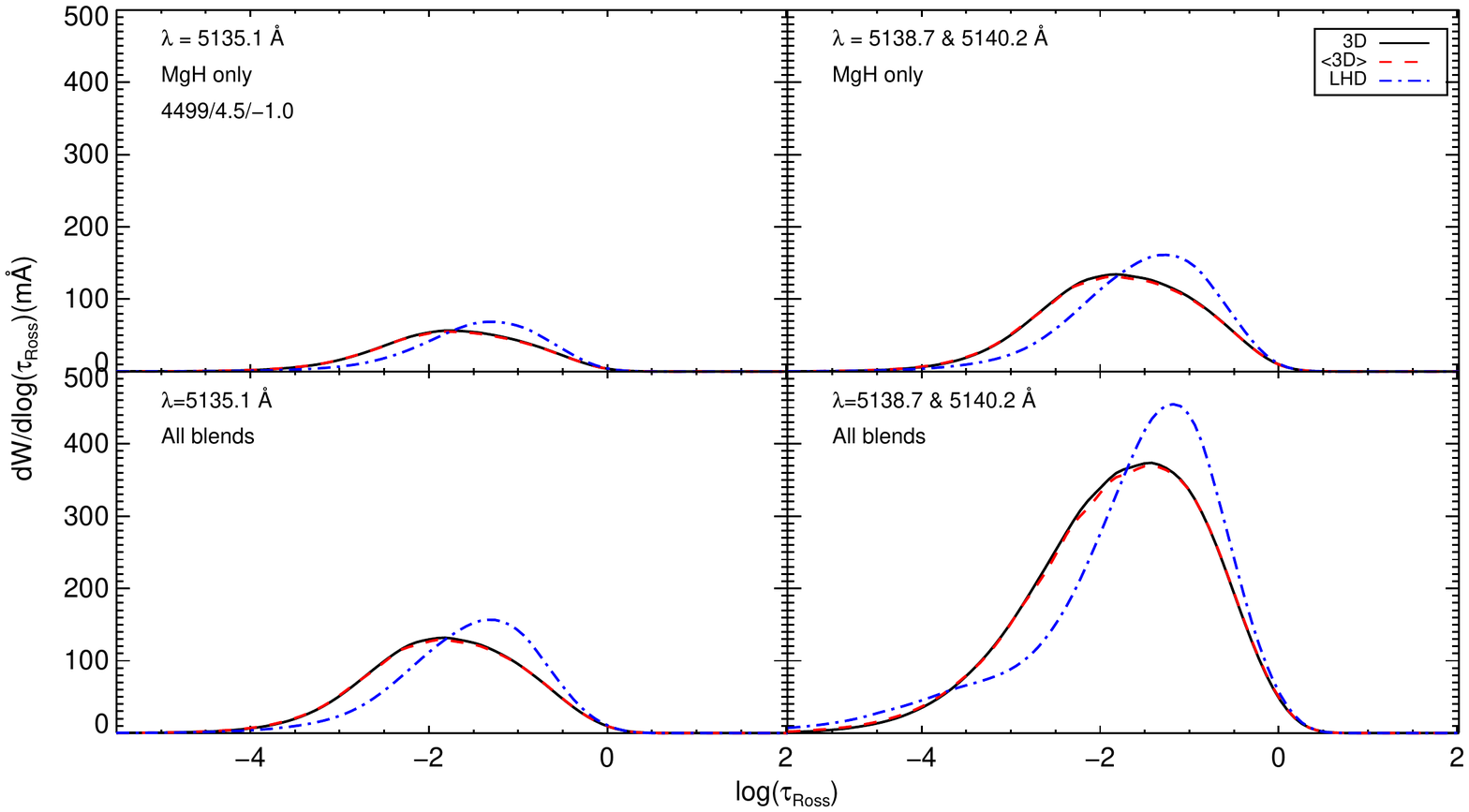}
\includegraphics[width=0.9\textwidth, trim = 2cm 4cm 0cm 4cm]{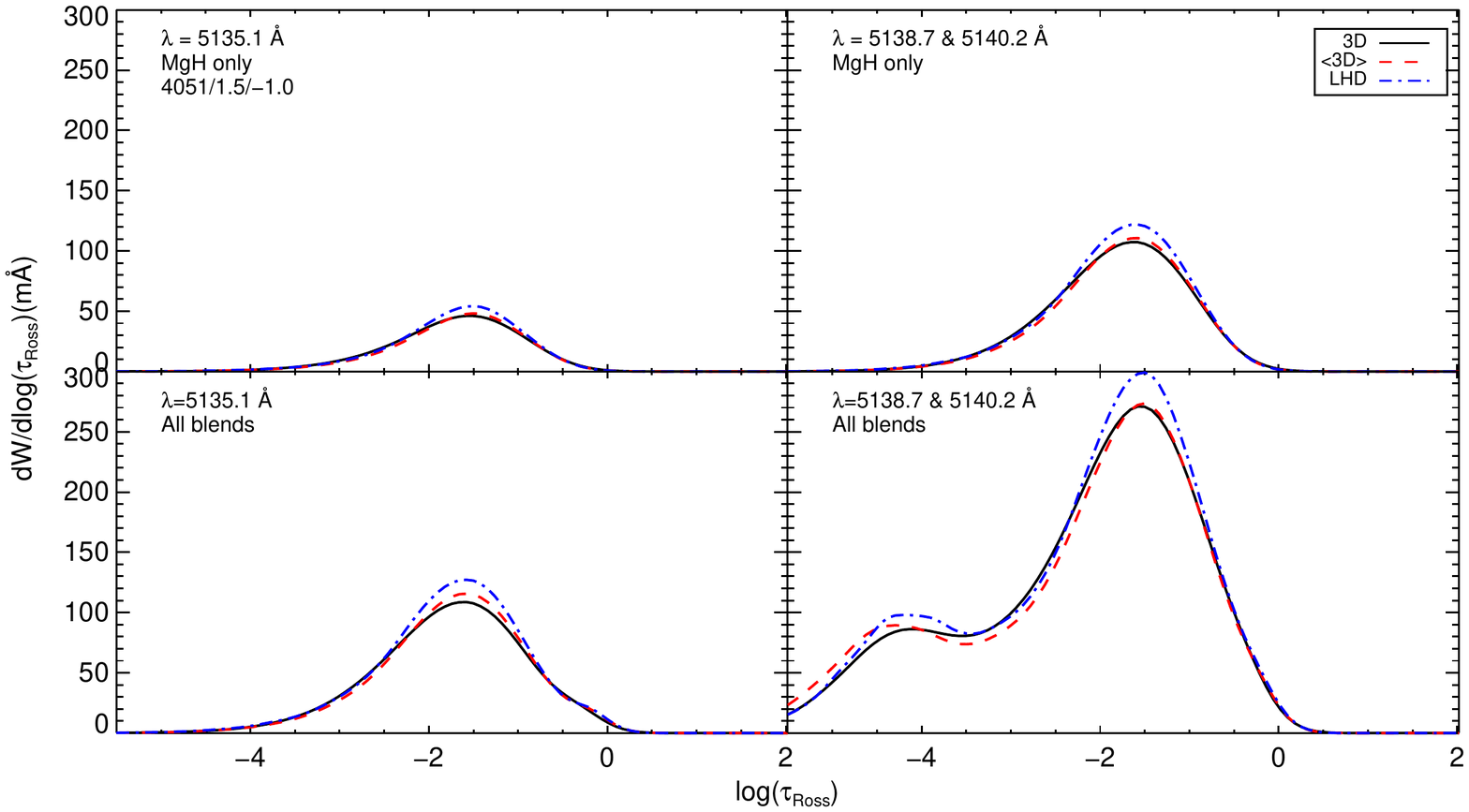}
\caption{Contribution functions for the region including the 5135\AA\ MgH feature and the region including the 5138\AA\ and 5140\AA\ features, for MgH only (top row for each model), and including all molecular and atomic blends (bottom row for each model). Top: The dwarf model d3t45g45mm10n01. Bottom: The giant model d3t4000g150m10x140z150k12.}%
\label{fig:contf-all}%
\end{figure*}

In reality, none of the MgH features modeled here are free from blends. Although many of the lines from the blending species are weak, they may react differently to 3D effects, resulting in differences in line shape and strength, that can affect the interpretation of observed spectra. In Fig.~\ref{fig:contf-all} we present contribution functions of the two regions computed for the purpose of the isotopic analysis for a typical dwarf (top) and a typical giant (bottom), including all blending lines. Also shown is the MgH only contribution functions for the 5135\AA\ and the entire region including the 5138.7~\AA\ and 5140.2~\AA\ features. The inclusion of blending lines clearly has a significant impact on the overall shape of the contribution functions, especially in the giant case.

Comparing the top rows of the dwarf and giant contribution functions in Fig.~\ref{fig:contf-all}, to the bottom rows in the same Figure, the 5135\AA\ region show similar behavior when all blends are included in both cases. The same holds for the region with the 5138.7~\AA\ and 5140.7~\AA\ MgH transitions for the dwarf. 

The giant model in Fig.~\ref{fig:contf-all} on the other hand, shows more pronounced differences when all blends are included. Here, the \avg\ synthesis shows a larger deviation from the full 3D case than when looking only at MgH transitions. This indicates that the temperature fluctuations are becoming more important for the line formation of the blends, than was the case for the dwarf model. This can be understood as a consequence of the larger $T$ fluctuations observed in giants, relative to the dwarfs, which will impact especially molecular lines \citep{gallagherCH}. The reason that the region including the 5138.7\AA\ and 5140.2\AA\ MgH features exhibits larger 3D-1D differences is a consequence of the broader wavelength range computed here, as this includes a larger number of blending species, compared to the 5135\AA\ region.

\subsection{Molecular number densities}
\begin{figure*}
\centering
\includegraphics[width=0.9\textwidth, trim = 4cm 4cm 4cm 3cm]{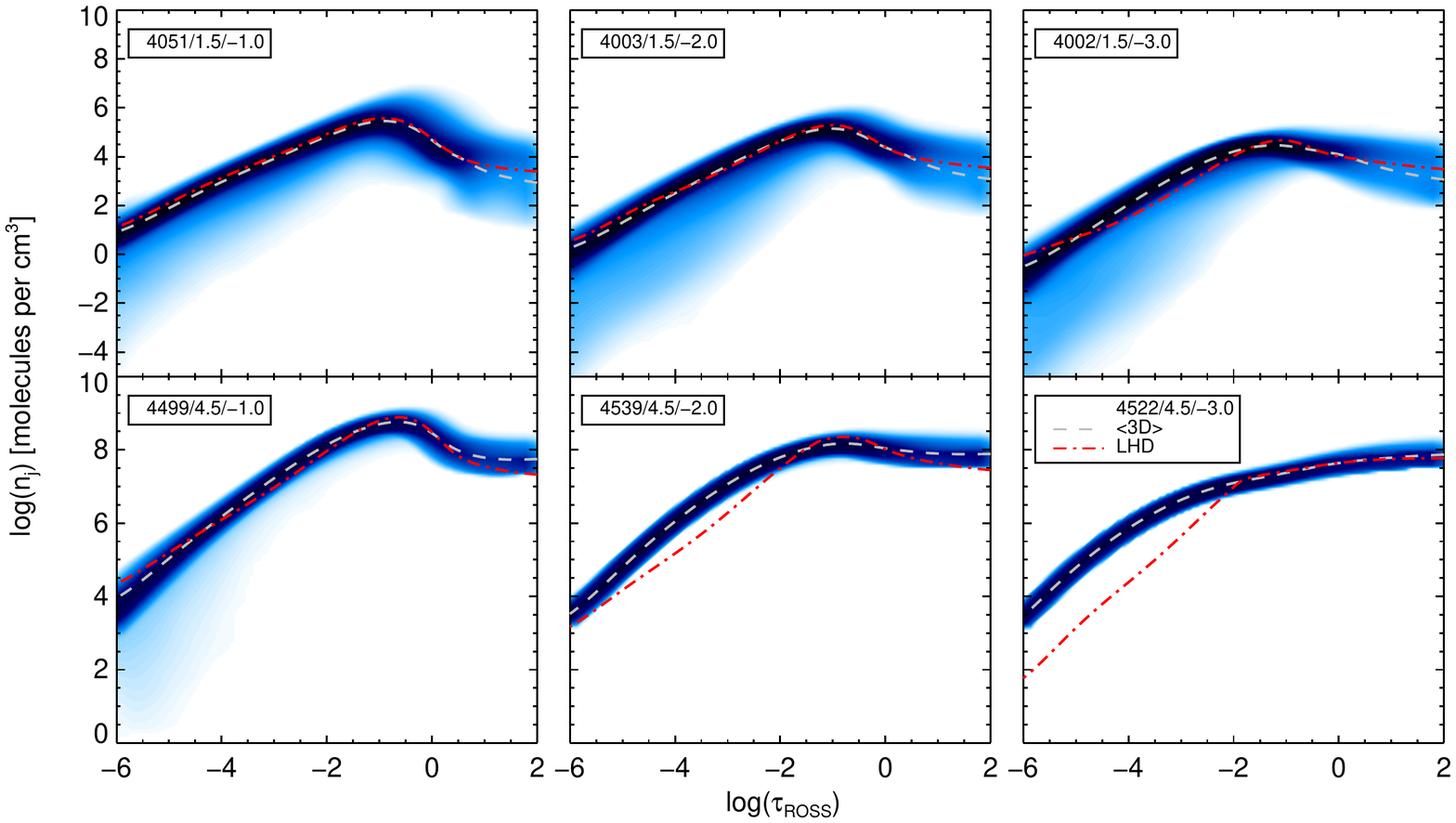}
\caption{Number densities of MgH for the 3D case (density plot), \avg\ case (white dash), and the 1D case (cyan dot-dash) for three giants (top) and three dwarfs (bottom) at different metallicities. Darker shaded regions indicate a higher sampling of points with this number density. See Table~\ref{tab:hydrogrid} for model details.}%
\label{fig:mgh-nd}%
\end{figure*}

The changes in the temperature structure of the model atmospheres, as well as the temperature fluctuations will impact the molecular number densities, $n_j$. Examples of this are shown in Fig.~\ref{fig:mgh-nd}, for three giants (top) and three dwarf (bottom) models, and were computed using the partition functions from \citet{1984ApJS...56..193S}. It is evident that convection impacts the formation of MgH to a larger extent in the giant case, where fluctuations in the MgH number density are observed in the outer parts of the atmosphere. This is significantly less pronounced in the dwarf case. However, both for giants and dwarfs, the number densities do not fluctuate as much as in the case for other molecules investigated in 3D (see \citealt{gallagherCH} for the dwarf case and \citealt{2007A&A...469..687C} for giants), and in all cases the results from the \avg\ models closely resemble the full 3D case. 

One possible explanation for this behavior could be the very low dissociation energy of MgH ($D_0=1.34$ eV), compared to CH, CO and OH investigated by \citet{gallagherCH} ($D_0=3.465, 11.092$ and $4.392$ eV respectively). As such, one might expect the number densities of the carbon bearing molecules to be more sensitive to $T$ fluctuations than MgH. To test this, we computed $n_j$ for CO and CH for the model d3t45g45mm30n01. It was found that the number densities of these molecules closely resembled what we found for MgH, with a narrow, symmetric distribution, without strong fluctuations. Thus, the main reason for the difference between the molecular $n_j$ in the dwarf models investigated here, compared to \citet{gallagherCH} must be related to the lack of $T$ fluctuations, and not to differences in $D_0$ between the molecules considered in the two cases. This was further confirmed by computing an experimental synthesis with $D_0$ of MgH artificially increased to 11 eV, which only served to increase $n_j$, but did not alter the shape of the distribution.

There is very little difference between the full 3D and the 1D results in the giant case, for all metallicities. The number densities are essentially the same, which is also supported by the MgH-only contribution functions, which show comparable strength and formation depth. 

For the metal poor dwarfs, this is no longer the case. While the temperature fluctuations due to convection have only a small impact, the significant changes to the overall temperature structure of the atmosphere between 3D and 1D models become important. Indeed, as seen in Fig.~\ref{fig:t-structure}, the temperature in parts of the line forming region is significantly cooler in the 3D/\avg\ case, compared to the 1D models, resulting in a significantly more efficient formation of MgH, by almost an order of magnitude in the main line forming region. Number densities for the models not shown in Fig.~\ref{fig:mgh-nd} are presented in Figs.~\ref{fig:all_nj_4k} and~\ref{fig:all_nj_5k} in Appendix~\ref{sec:nj}.

\section{Analysis of 3D MgH features}\label{sec:analysis}

\subsection{Micro and macroturbulence}
The self-consistent modeling of convection in 3D hydrodynamical model atmospheres has several advantages over 1D models. In particular the models aim at a realistic modeling of the turbulent gas motion, so that it is no longer required to parametrise this using microturbulence, \vmic, and macroturbulence, \vmacro. However, since the 3D gas motions are undoubtedly impacting the line shapes, this needs to be accounted for in the 1D syntheses before a meaningful comparison can be made between 1D and 3D. This is especially important when dealing with isotopic ratios, as the line broadening applied in 1D will have a significant impact on the line shapes and hence the derived abundance of the Mg isotopes.

To derive the \vmic\ and \vmacro\ of the 3D models we used four different lines, three Ca-I lines at 5581\AA, 5590\AA, and 5601\AA, and one Ti-I line at 5145\AA.  These lines all have 1D contribution functions very similar to those of MgH, and have been used in previous studies of Mg isotopes to derive \vmacro\ in 1D syntheses (\citealt{yongiso2003} and \citealt{melendez07}).

We utilized method 2a from \citet{steffenvmic} to derive \vmic. In brief, from each of the atomic transitions we generated a set of nine artificial lines, changing the \loggf\ value to get a variation in line strength, but kept all other line parameters intact. A full 3D synthesis was computed for these sets, constituting 36 lines in total. In addition, we computed a set of 1D syntheses for each line, varying \vmic, but also again changing the \loggf\ for each line. This provided a set of 1D synthesis with different \vmic\ and \loggf\ for each of the 36 lines computed in full 3D. We changed the \loggf\ values to account for possible differences in line strength between 1D and 3D. We assumed that changing the \loggf\ value was equivalent to changing the abundance, $\log\epsilon$. The advantage of changing \loggf\ over changing the raw abundance of the lines in question is that we do not need to consider possible changes to the free electron budget. This could in principle impact the continuum opacity through H$^-$. By changing the \loggf\ of the lines instead, we circumvented this.
 
For each 3D line equivalent width, $EW_{\mathrm{3D}}$, we thus have a set of $EW_{\mathrm{1D}}(\xi_{\mathrm{micro},j},\log\epsilon_{\mathrm{1D},i})$. For each $\xi_{\mathrm{micro}}$, we then computed the abundance correction, $\Delta\log\epsilon_{\mathrm{1D},i}(\xi_\mathrm{micro})$, so that $EW_{\mathrm{3D}}=EW_{\mathrm{1D}}$. To finally derive the microturbulence in the 3D model we performed linear fits to $\{EW_{\mathrm{3D}},\Delta\log\epsilon_{\mathrm{1D}}(\xi_\mathrm{micro})\}$ and demanded that the relation had a slope of zero. We ensured that only lines with an EW less than 150m\AA\ were used for this exercise, to avoid problems with strong saturation.

\begin{figure*}
\centering
\includegraphics[width=0.9\textwidth, trim = 4cm 6.5cm 4cm 7cm]{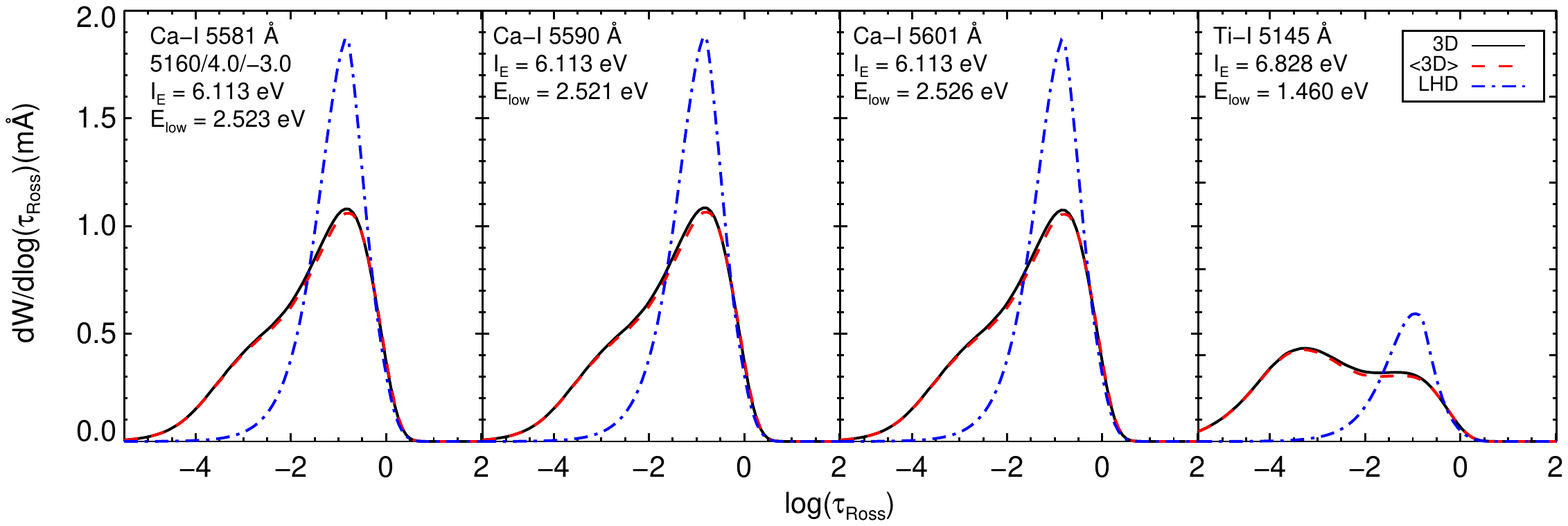}
\caption{Contribution functions for the lines used to determine \vmic\ and \vmacro\ for model d3t50g40mm30n02.}%
\label{fig:cati-contf}%
\end{figure*}

The three Ca-I lines gave consistent values for \vmic\ across all 3D models, while the \vmic\ derived from the Ti-I line showed some deviation. We attributed this to the fact that while the selected elements have comparable ionization energy (Ca-I = 6.11 eV, Ti-I = 6.83 eV; \citealt{NIST_ASD}), the excitation potentials differ by $\sim1$~eV. Since the line strength depends on both temperature as well as these two energies, the Ca and Ti lines will have different sensitivity to the temperature fluctuations and the different temperature structure in the 3D models. Indeed, for the low-metallicity dwarfs, the contribution functions for the Ca lines and Ti lines can be very different between 1D and 3D (Fig.~\ref{fig:cati-contf}), with the 3D lines forming over a much larger range of optical depths than the 1D case. As such it is not surprising that we cannot describe \vmic\ with a single value for both the Ca and Ti lines. The contribution functions of the Ca lines were more similar to the MgH contribution functions than the single Ti line, and we discarded the latter in the determination of \vmic\ and \vmacro. As our final value we took the average of the individual Ca lines. We report these in Table~\ref{tab:hydrogrid}.

While the \vmic\ determination gives reasonable, albeit low, values in most cases, it is worth noting that there seems to be a correlation between \vmic\ and metallicity, with the \vmic\ increasing as the metallicity decreases. A similar effect was seen by \citet{steffenvmic}, who attributed this not to the low-metallicity models having an abnormally high \vmic\ but rather an effect of the changes in temperature structure between 3D and 1D models. The metal-poor 3D models are significantly cooler in the line forming region than their 1D counterparts. Thus strong, low excitation potential lines will require a higher abundance correction than weaker lines, to match the 3D EW. Since we determine the \vmic\ by requiring identical abundance corrections across all line strength, this will tend to increase \vmic\ as a high value of \vmic\ will compensate for a line-dependent abundance correction.

One model, d3t40g45mm20n01, stands out, giving a microturbulence of zero. While this might seem surprising, an inspection of the temperature structure revealed that the 3D model has very efficient convection, with essentially adiabatic stratification out to an optical depth of $\sim-2.0$, also for the 1D model (Fig.~\ref{fig:allTtau_4k}). The effective temperature of the model together with efficient convection implies small convective velocities and small temperature fluctuations, so that thermal motion will dominate the line broadening. Moreover, the $T$-distribution is very narrow and practically identical between 3D and 1D. As such, one would not expect weak lines forming in this region to require any significant \vmic, nor abundance corrections to compensate for line strength differences, as they would be near-identical already.

Before performing any additional analysis of the 3D spectra, we convolved all spectra with a Gaussian with a full-width at half-maximum (FWHM) corresponding to $R=110\,000$, which is the typical resolving power of an UVES-like spectrograph used for observing stars for the purpose of studying MgH isotopes.

With the 3D model \vmic\ determined, we derived the \vmacro\ by fitting the same 3D spectral lines with 1D syntheses computed with the \vmic\ and $\Delta\log\epsilon_{\mathrm{1D}}$ just determined, but a range of broadenings. We assumed that the \vmacro\ was isotropic and described by a Gaussian, and took the best-fitting value as determined by a $\chi^2$ minimum, to represent the macroscopic turbulence in our 1D syntheses. The value of \vmacro\ reported in Table~\ref{tab:hydrogrid} corresponds to the straight average of the values from the individual lines. Before fitting the 1D syntheses, we performed a wavelength shift to account for the convective line shift present in the 3D syntheses. 

As can be seen from Table~\ref{tab:hydrogrid}, we found that the giant models have larger \vmacro\ than the dwarfs. It is also worth noting that the determined \vmacro\ decreases with decreasing metallicity, for both dwarfs and giants. This can be attributed to the increased \vmic, which will already introduce additional line broadening as discussed above. In effect that results in a small (or none) subsequent macroturbulent broadening needed in the 1D syntheses to provide the best match to the 3D syntheses, effectively creating an anti-correlation between \vmic\ and \vmacro.

\subsection{3D-1D corrections}
\label{sec:1D-3Dcorr}
In Fig.~\ref{fig:3D-synth-novmic} we show \mghi-only syntheses for all three features for our coolest dwarf model for $[\alpha/\mathrm{Fe}]=+0.4$. Both the original 3D and best-fitting 1D syntheses are shown, and we indicate the central wavelength of the \mghi\ isotopic component for each feature. The best fits were determined through $\chi^2$ minimization, varying the abundance of Mg, as well as the fraction of \mgii\ and \mgiii, under the constraint that the total amount \mgi+\mgii+\mgiii=100. In addition, a small wavelength shift was permitted in order to account for possible convective velocity shifts. Even with \vmic\ and \vmacro\ values of 0 \kms\ and a reduced [Mg/Fe], the 1D syntheses are still broader than the full 3D syntheses, particularly evident for the 5138\AA\ feature. In addition, the 1D syntheses are not able to simultaneously reproduce the 3D line shape and the core strength.

To investigate whether the 1D dwarf syntheses were systematically broader than the 3D counterparts we analysed the \mghi\ only syntheses for all our dwarf models. The 1D and \avg\ syntheses were computed with identical values of \vmic. We performed Gaussian fits to the 1D, \avg\ and 3D syntheses for each model and computed the FWHM, which we used to quantify the total line broadening. In essentially all cases it was found that $FWHM_\mathrm{1D}>FWHM_\mathrm{3D}\geq FWHM_{\langle\mathrm{3D}\rangle}$. Only for the two hottest models at $[$Fe/H$]=-1.0$, were the 3D syntheses broader than both the \avg\ and 1D syntheses. This is consistent with these models also having the largest $T$-fluctuations amongst the dwarfs (Fig.~\ref{fig:allTtau_5k}). That the 1D syntheses are broader than the \avg\ models, suggests that the main reason for the discrepancy is the difference in the overall temperature structure between 3D and 1D. The line forming regions in 1D are significantly hotter, so the thermal broadening of the lines will be larger. This is true even for the coolest dwarfs which show significant differences in their structure between 3D and 1D, already for a metallicity of $-1.0$ (Fig.~\ref{fig:allTtau_4k}). We note that these syntheses were computed \textit{without} applying any macroturbulent/instrumental broadening, although that would not have changed the result.

Two things can be learned from this:

\begin{figure*}
\centering
\includegraphics[width = 0.9\textwidth, trim = 3cm 5cm 3cm 5.5cm]{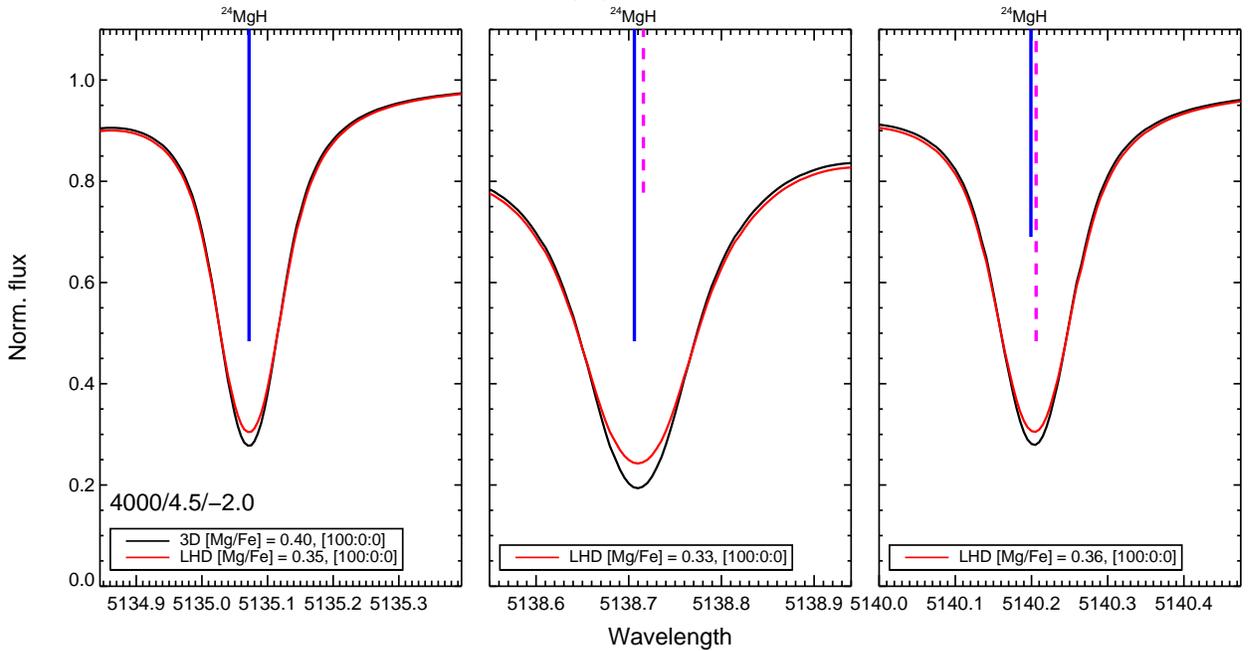}
\caption{Syntheses of the three MgH features in the dwarf model d3t40g45mm20n01. 3D (solid, black) and 1D (solid red) syntheses are shown. The best-fitting values of the Mg abundance and \mgi:\mgii:\mgiii\ 1D fit are shown in the legend. This synthesis included only \mghi. We indicate the central wavelength of the \mghi\ features with vertical blue and magenta dashed lines, scaled relative to the strength of the strongest component of the 3D syntheses in each panel.}%
\label{fig:3D-synth-novmic}%
\end{figure*}

First, the difference in line strength means that any 1D fit will tend to overestimate the abundance of Mg, as this will need to be artificially increased to match the full line strength. This was already observed by \citet{ramirez} in the case of the solar metallicity dwarf, HIP 86400 (\teff\ $=4830$ K, \logg\ $=4.62$, \feh\ $=-0.05$). They found a significantly better match between observations and the spectral synthesis from a 3D model, compared to that from a traditional 1D model, when computed with identical Mg abundance. 

Second, the broader lines found in the 1D case will tend to result in an underestimation of the amount of the heavy isotopes, when fitting spectra with syntheses based on 1D model atmospheres.

To investigate this in more detail, we performed fits with 1D syntheses to a number of 3D syntheses, using the \texttt{Fitprofile} software described in \citet{thygesen}. We simultaneously fit the abundance of magnesium, as well as the fraction of \mgii\ and \mgiii. The fraction of \mgi\ was subsequently computed as $100-^{25}\mathrm{Mg}-^{26}\mathrm{Mg}$. \texttt{Fitprofile} also allows for small velocity shifts when performing the $\chi^2$ minimization, to account for convective line shifts not present in the 1D models. 

For each 3D model we analyzed three different syntheses, with isotopic compositions 100:0:0, 94:3:3 and 88:6:6 percentage \mgi:\mgii:\mgiii. This covers the expected range of isotopic compositions in stars at these metallicities \citep{yongiso2003}. All syntheses were computed with $[\alpha/\mathrm{Fe}]=+0.4$. 

\begin{figure*}
\centering
\includegraphics[width = 0.9\textwidth, trim = 3cm 4.7cm 3cm 5cm]{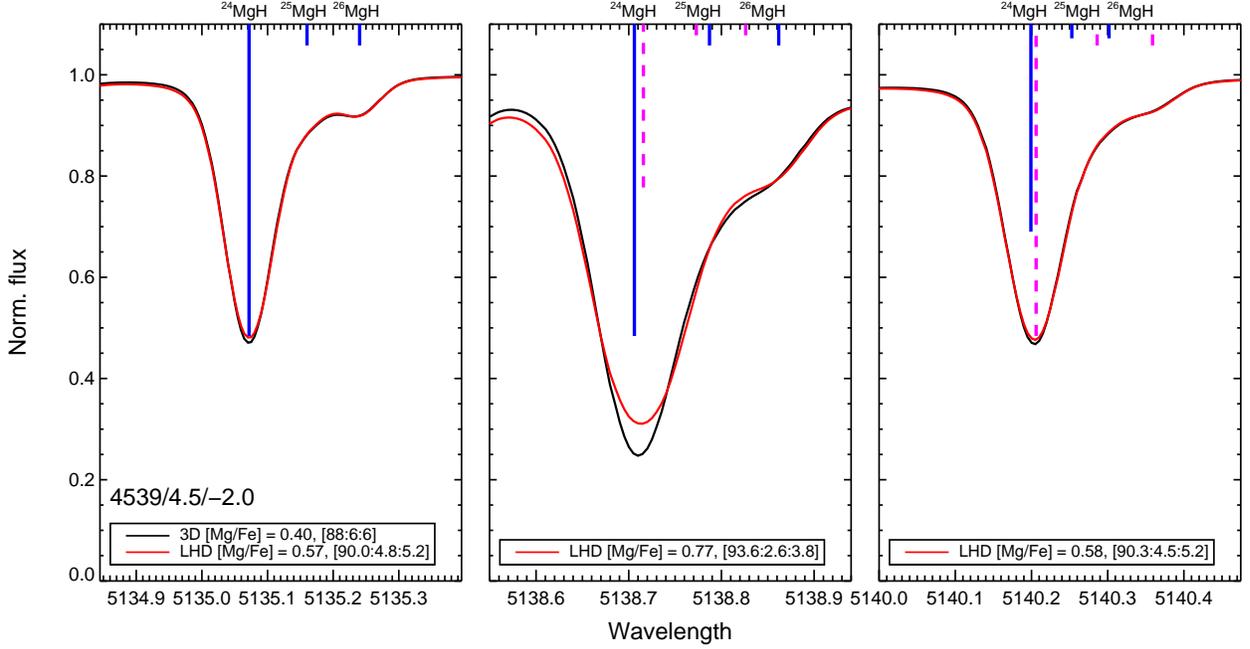}
\caption{1D fit (red) to the d3t45g45mm20 3D syntheses for each of the three features. Original composition (3D) and best fit values (1D) are provided in the captions. The central wavelength of the MgH features are shown as vertical solid blue and dot-dashed magenta lines. The length of the lines are scaled relative to the strongest component in each panel, and to isotopic fraction of the 3D syntheses.}%
\label{fig:1D-3Dfit}%
\end{figure*}

Figure~\ref{fig:1D-3Dfit} shows the best-fitting 1D syntheses for each of the MgH features for a typical dwarf model, with the 3D syntheses computed for an 88:6:6 isotope composition. The 1D syntheses are reproducing the line shapes well for the 5135\AA\ and 5140\AA\ features in this case, although the tendency of the 1D syntheses to be slightly too broad is still visible. The situation is markedly different for the 5138\AA\ feature, where neither the line shape nor the line strength is well reproduced in 1D, even considering the substantial enhancement in Mg. The fraction of the heavy isotopes are also overestimated for all three features. This confirms the expectations from Fig.~\ref{fig:3D-synth-novmic}, that the 1D syntheses would tend to yield lower amounts of the heavy isotopes, even if we are able to reproduce the shape of the MgH wings well. The problems with reproducing the MgH feature at 5138\AA\ is persistent across most dwarfs models, for the three isotopic mixtures considered here, although we note that the agreement improves for the 5000K models.

\begin{figure*}
\centering
\includegraphics[width = 0.9\textwidth, trim = 3cm 4.7cm 3cm 5cm]{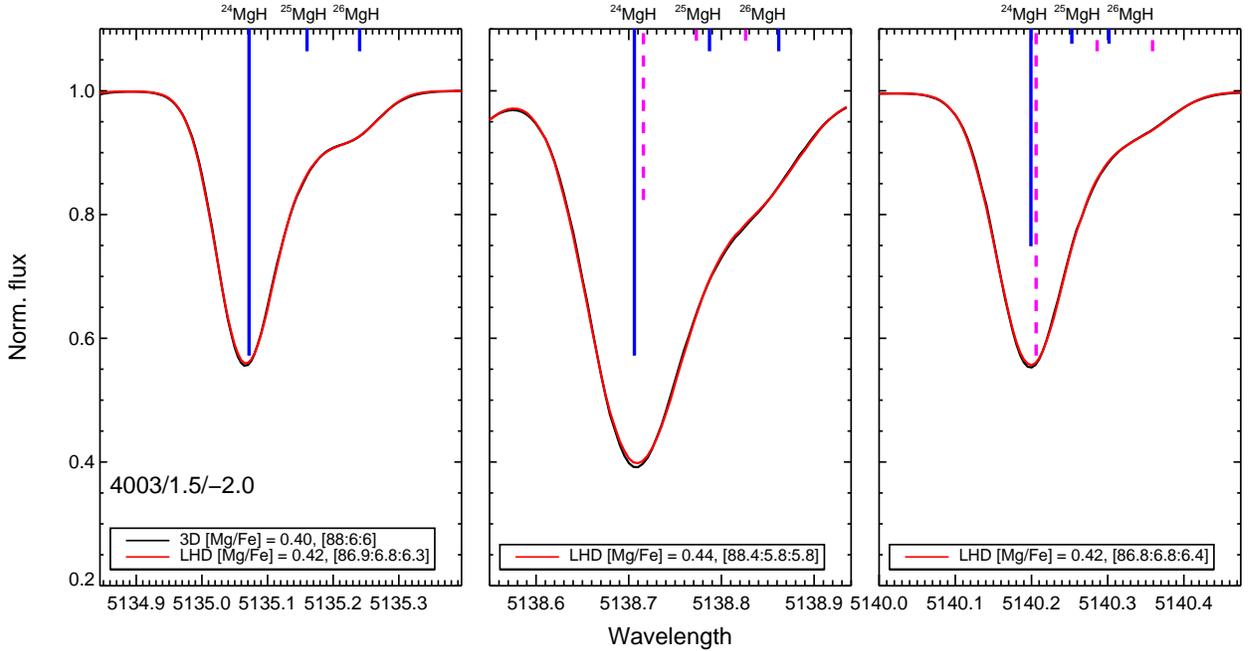}
\caption{1D fit (red) to the d3t4000g150m20x140z150k14 3D syntheses for each of the three features. Original composition (3D) and best fit values (1D) are provided in the captions. The central wavelength of the MgH features are shown as vertical solid blue and dot-dashed magenta lines. The length of the lines are scaled relative to the strongest component in each panel, and to isotopic fraction of the 3D syntheses.}%
\label{fig:1D-3Dfit_giant}%
\end{figure*}

In the case of a giant model, the 1D syntheses were also able to fit the 3D spectra well (Fig.~\ref{fig:1D-3Dfit_giant}). Both the line strengths and the overall line shapes were reproduced remarkably well in the 1D case for this particular model, with the \mgii\ fraction being slightly overestimated. The best-fitting Mg abundances were essentially identical to the input abundance. The good reproduction of the 3D line profile remains for the giant models at $[\mathrm{Fe}/\mathrm{H}]=-1.0$ and $-3.0$, but in these cases the disagreement between the input Mg abundance and the best-fitting value increased. Nevertheless the agreement between 1D and 3D for the Mg abundance was still significantly better than for the dwarfs, at most differing by 0.16 dex. The isotopic fractions, on the other hand, stays essentially unchanged, differing at most by $1.4$ percentage points for \mgi\ in the case of the 88:6:6 mixture for the $[$Fe/H$]=-3.0$ metallicity giant. The differences for the heavy isotopes across all remaining features and giant models were typically $\leq1$ percentage point.

In Table~\ref{tab:3dcorrections} we provide the 3D corrections to the derived Mg abundances, in the sense $\Delta$Mg$_\mathrm{3D-1D} = A(\mathrm{Mg})_\mathrm{3D}-A(\mathrm{Mg})_\mathrm{1D}$. We averaged the results from the fits of the individual features, and note that the feature-to-feature agreement is typically good, differing only by a few hundredths of a dex in the vast majority of the cases.

\begin{figure*}
\centering
\includegraphics[width = 0.9\textwidth]{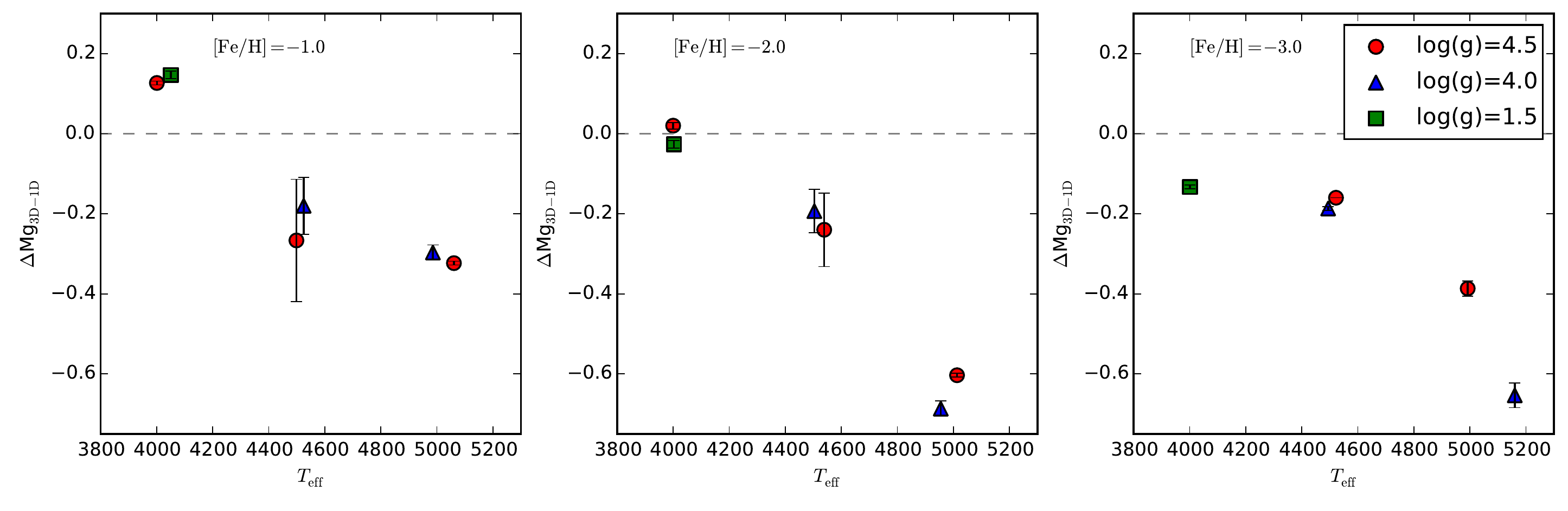}
\caption{$\Delta$Mg$_\mathrm{3D-1D}$ for all models, against \teff. Model gravities are: \logg=1.5 (green squares), \logg=4.0 (blue triangles), and \logg=4.5 (red circles). Each panel is for a fixed metallicity. The errorbars give the standard deviation of the mean, and are smaller than the size of the plotting symbols in most cases.}%
\label{fig:3Dcorrects}%
\end{figure*}

In Fig.~\ref{fig:3Dcorrects}, we plot the $\Delta$Mg$_\mathrm{3D-1D}$ value for all models in our sample, including errorbars given as the standard deviation. It is evident that there is a correlation with \teff\, with hotter models showing a larger deviation in Mg abundance, peaking at $[\mathrm{Fe/H}]=-2.0$ for the 5000K, \logg=4.0 model. It is clear that also the metallicity has an impact, both for giants and dwarfs, where the correction changes sign for the giants as the metallicity decreases. The hottest models also exhibit larger corrections as the metallicity drops below $[\mathrm{Fe/H}]=-1.0$. The 4500K dwarfs, on the other hand, appears to have a nearly constant abundance offset across all metallicities. \citet{2004ApJ...603..697Y} studied Mg isotopes in a large sample of Hyades dwarfs, and also found a correlation between the Mg abundance derived from the MgH features, and the stellar \teff, with stars at 5000K having $\sim0.2$ dex higher Mg values than their counterparts at 4000K. They did, however, find that the Mg abundance derived from the MgH features were smaller than the equivalent derived from atomic lines. This, together with the smaller range found by these authors is likely a consequence of the metallicity of the Hyades being slightly above solar. This is supported by our model comparisons, that suggest that the total range of Mg corrections decrease, and that the $\Delta$Mg$_\mathrm{3D-1D}$ may change sign as the metallicity increases as indicated by the correction for the 4000K dwarf at $[$Fe/H$]=-1.0$.

The contribution functions shown in Figs. \ref{fig:contf-mgh-dwarf} and \ref{fig:contf-all} suggest that the MgH line equivalent widths are relatively similar between 1D and 3D, especially for the most metal rich dwarf shown. That we still found a need for abundance corrections when fitting our 3D syntheses with 1D syntheses is a consequence of the \mghi\ component already dominating the spectrum. For the heavy isotope fractions considered here, any decrease(increase) in the amount of \mgii\ and \mgiii\ will also decrease(increase) the total line EW, as the \mghi\ component barely changes strength, since the vast majority of the available Mg is already in this form. To match the line asymmetries it was found that a decrease in the heavy isotopes was required in the 1D syntheses (see below). This, in turn, allowed for a better match to the 3D core line strength by increasing the Mg abundance. As such, not only the core strength, but also the line shape affects the derived 3D$-$1D Mg abundance corrections.

\begin{table}
\centering
\caption{The 3D-1D corrections for the Mg abundance, derived from $\chi^2$ fits to the 3D syntheses of the three MgH features. \label{tab:3dcorrections} }
\begin{tabular}{lc}
\hline \hline
Model & $\Delta$Mg$_\mathrm{3D-1D} $ \\
\hline
d3t4000g150m10x140z150k14 &  $+0.15$ \\
d3t4000g150m20x140z150k14 &  $-0.03$ \\
d3t4000g150m30x140z150k14 &  $-0.13$ \\
\hline
d3t40g45mm10n01 &  $+0.13$ \\
d3t40g45mm20n01 &  $+0.02$ \\
d3t45g40mm10n01 &  $-0.18$ \\
d3t45g40mm20n01 &  $-0.19$ \\
d3t45g40mm30n02 &  $-0.19$ \\
d3t45g45mm10n01 &  $-0.27$ \\
d3t45g45mm20n01 &  $-0.24$ \\
d3t45g45mm30n01 &  $-0.16$ \\
d3t50g40mm10n01 &  $-0.30$ \\
d3t50g40mm20n01 &  $-0.69$ \\
d3t50g40mm30n02 &  $-0.65$ \\
d3t50g45mm10n03 &  $-0.32$ \\
d3t50g45mm20n03 &  $-0.60$ \\
d3t50g45mm30n03 &  $-0.39$ \\
\hline
\end{tabular}
\end{table}

\begin{figure*}
\centering
\includegraphics[width = 0.9\textwidth]{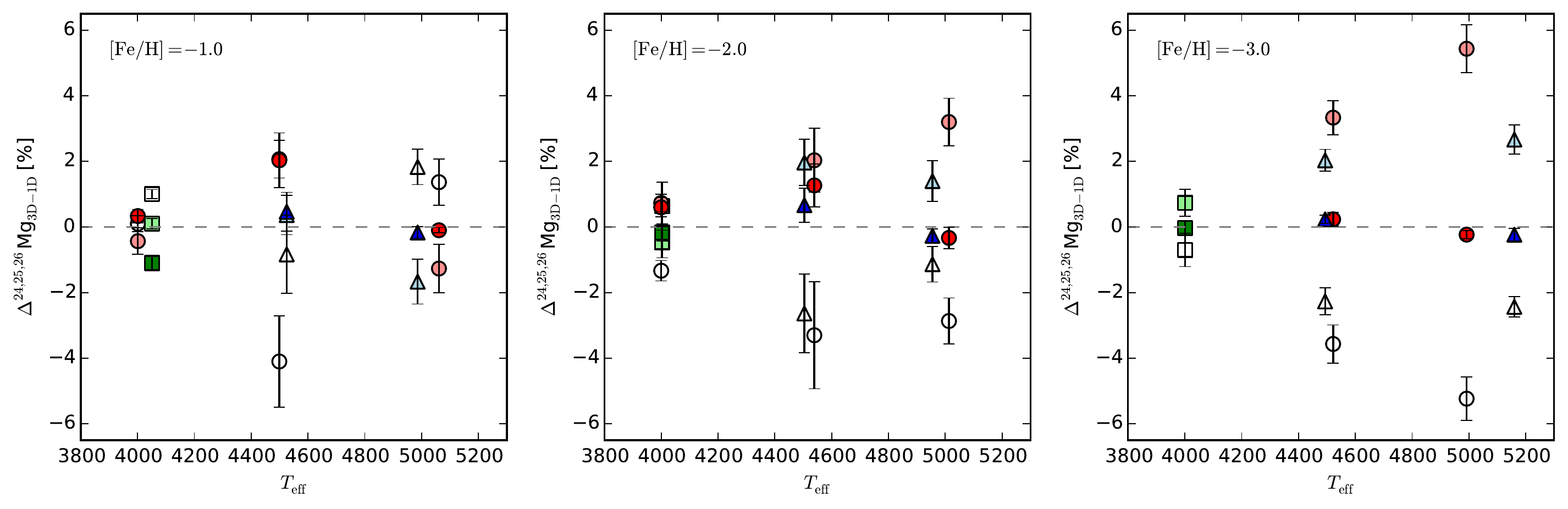}
\caption{$\Delta^{24,25,26}\mathrm{Mg}_\mathrm{3D-1D}$ fractions for all models, against \teff. Symbol shapes have same meaning as in Fig.~\ref{fig:3Dcorrects}, but here are color coded according to the specific isotope (\mgi, open symbols), (\mgii, light shaded symbols), (\mgiii, dark shaded symbols). The errorbars give the standard deviation of the mean.}%
\label{fig:MgHcorr}%
\end{figure*}

Deriving 3D-1D corrections for the isotopic fractions of Mg was more subtle, as this required an accurate reproduction of the full line shape, which was not possible for the 5138\AA\ feature in the dwarf models. Nevertheless, it is instructive to inspect the differences between 1D and 3D also for the isotopes, to assess qualitatively, whether any systematic effects are present. In Fig.~\ref{fig:MgHcorr} we plot $\Delta^{24,25,26}$Mg$_\mathrm{3D-1D}$ fractions for all our models, based on the syntheses with the 88:6:6 isotopic composition. While these differences should not be considered as absolute corrections to be applied to 1D analyses of real stellar spectra, some systematic behavior is still evident. 

The magnitude of the differences between 3D and 1D increases for the dwarf models as the metallicity decreases. We attribute this to the changes in the overall temperature structure of the model atmospheres, which increases when the metallicity decreases as discussed in Sect.~\ref{sec:modeling}. As a result, lines of MgH, as well as blending species will form in layers of the atmosphere with significantly different temperatures, which impacts the line formation. It also appears that the 1D models tend to underestimate the \mgii\ fraction and overestimate the amount of \mgi, relative to the input composition in the 3D syntheses. The fraction of \mgiii, on the other hand, is recovered well in most cases. This is a consequence of the different shape of the \mghi\ feature in 3D and 1D, which will have a significantly stronger impact on the neighboring \mghii\ feature than on the \mghiii\ feature. This effect is also seen for the two 5000K models at  $[\mathrm{Fe/H}]=1.0$, but in the opposite direction, where \mgi\ is underestimated and \mgii\ is overestimated, consistent with our finding that for these two models, the 3D syntheses were broader than their 1D equivalents.

It is also clear that the disagreement between dwarfs in 3D and 1D is correlated with the stellar \teff\ for the \logg $=4.5$ models, for a fixed metallicity, when the metallicity is below $-1.0$. Albeit small, the differences more than double when increasing the model \teff\ from 4000K to 5000K for \feh\,$=-2.0$.

The correlation between metallicity and 3D-1D differences for the dwarfs may also be related to the high value of \vmic\ found for the low metallicity models. If this is overbroadening the 1D syntheses, the best fit will tend to yield a lower fraction of the heavier isotopes. However, inspecting Fig.~\ref{fig:3D-synth-novmic} it is clear that even for a model where we derived both a \vmic\ and \vmacro\ of zero \kms\ from fitting the Ca lines, the 1D syntheses are still slightly broader than the 3D equivalent. Keeping in mind the impact of the thermal broadening discussed above, we consider it unlikely that the high value of \vmic\ for the lowest metallicity is the explanation for the disagreement. Rather the differences in $T$ structure between 3D and 1D models create the observed effect.

The giants, on the other hand, show a somewhat different behaviour. Here, the disagreement between 3D and 1D is small for all metallicities, with a tendency to slightly overestimate the amount of \mgii\ and \mgiii\ in 1D, which naturally leads to an underestimation of \mgi. Since we only have giant models at a single temperature and \logg, we cannot comment on possible systematic behaviour of the isotopes with these two parameters. 

We note that the same behavior was also observed in the case where the fractions of heavy isotopes of Mg were halved, compared to the case discussed here, \ie\ a 94:3:3 mixture. For the case of the pure \mgi\ mixture, the majority of the best fitting dwarf 1D syntheses agree with the input isotopic values. The only exception being the two 5000K models at $[\mathrm{Fe/H}]=-1.0$, where a slight overestimation of \mgii\ was found (at most 2.4 percentage points for the 5140\AA\ feature). We note that the good agreement with the \mgi\ only mixture should be taken with some caution, since the 1D syntheses are broader than their 3D counterparts. Our fitting routines do not allow for negative isotopic fractions, which undoubtedly would provide a better agreement in a $\chi^2$ sense. For the giants, on the other hand, the best fit to the pure \mgi\ syntheses resulted in a marginal overestimation ($<1$ percentage point) of the heavy isotopes, consistent with our findings from the two isotopic mixtures discussed above. 

\section{Discussion}\label{sec:discussion}

From the investigation performed here, it is evident that effects from using 3D hydrodynamical atmospheres in place of traditional 1D atmospheric models impacts the lines from MgH in a non-negligible fashion. This is particularly true for dwarf stars, where the differences in the $T(\tau)$ structure between 1D and 3D are more dramatic than for the giants. Especially the core strength of the MgH lines changes significantly between 1D and 3D, a fact that was already noted by \citet{ramirez}, with the 3D synthesis providing a significantly better fit to the observed stellar spectra. The magnitude of the $\Delta$Mg$_\mathrm{3D-1D}$ corrections increases with increasing model \teff, but also metallicity changes have an impact. The sensitivity to metallicity changes can be attributed to changes in the overall $T(\tau)$ structure, where the 1D models will tend to be significantly hotter than their 3D counterparts in the line forming region. The cooler 3D atmospheres will enhance the molecular formation, allowing the lines to form over a more extended part of the model atmosphere. In 1D, on the other hand, the hotter atmospheres means that the lines form in deeper layers, where the pressure is high enough for MgH to form in a significant amount. The increased temperature results in increased thermal broadening, as well as stronger molecular dissociation. The combination of these effects results in a need for a Mg enhancement in the 1D syntheses when simultaneously fitting the line shape and line strength of the 3D synthesis.

In fitting of the MgH lines in observed spectra, the Mg abundance is normally included as a fitting parameter in order to match the line strength of the MgH features (e.g. \citealt{yongiso2003} and \citealt{melendez07}). In most cases this results in a disagreement between the Mg abundance found from the atomic lines in the stellar spectra, and that needed to fit the MgH features. The results from Sect.~\ref{sec:1D-3Dcorr} suggests that 1D dwarf syntheses will tend to overestimate the Mg abundance at low metallicity, compared to the 3D case. \citet{2004ApJ...603..697Y} already reported correlations between stellar parameters and [Mg/Fe] from the MgH features for the Hyades. The behavior of the Mg abundance derived from the molecular features at low metallicity dwarfs is currently under investigation.

Since an accurate determination of the Mg isotopes in stars must necessarily demand a good reproduction of the entire line profile, not merely the wings, any analysis using 3D atmospheres will impact the observed isotopic ratios. The results from Sect.~\ref{sec:1D-3Dcorr} suggests that in the case of low metallicity dwarfs, previous analyses relying on 1D models are likely to have underestimated the fraction of \mgii\ observed in stellar spectra by a few percentage points, while the fraction of \mgiii\ appears to be largely unchanged. From an observational point of view, this is fortunate, as the \mgiii\ isotope is the easiest to measure and is considered more reliable than the \mgii\ measurements, due to the larger wavelength separation from the strong \mghi\ lines, that dominate the MgH features in all cases. 

The potential increase in \mgii\ at the lowest metallicities may bring observations at odds with current chemical evolution models for the Milky Way halo (e.g. \citealt{kobayashi}), which predicts a very low \mgii/\mgi\ ratio at metallicities below $-1.5$. This in turn may require modifications of the chemical evolution models for the halo, but before the impact can be determined, a large sample of halo stars needs to be analyzed using syntheses based on 3D model atmospheres. Such a project is currently underway for a substantial sample of metal-poor halo stars. 

\section{Conclusion}\label{sec:conclusions}

In this paper we have presented the first ever detailed analysis of the formation and shape of optical MgH lines, using 3D hydrodynamical \cobold\ stellar atmospheres in the context of isotopic ratios. Based on the comparison between 3D and 1D syntheses we found that the Mg abundance required to fit the 3D features, in most cases needed to be enhanced beyond the nominal input value of $[\alpha/\mathrm{Fe}]=+0.40$. This was especially true for dwarf stars, where an additional enhancement of up to $+0.69$ dex was required for the 5000K, 4.0 \logg\ model at $[\mathrm{Fe/H}]=-2.0$, clearly outside any reasonable value. The disagreement between 1D and 3D Mg abundances was found to increase with increasing temperature, but also the metallicity plays a role, likely as an effect of changes to the overall $T(\tau)$ structure being significantly different in 1D and 3D. 

The influence of 3D atmospheres on the isotopic fractions are less dramatic than the overall line strength. For giants the 1D models reproduce the full line shape well, and the isotopic fractions are well recovered, with the largest differences being approximately one percentage point, for the heavy isotopes, which is comparable to the best-case fitting precision when dealing with high-quality observed spectra. 
For dwarfs the disagreement is more substantial. We were able to simultaneously fit the core strength and the wings of the MgH lines at 5135\AA\ and 5140\AA, but the 1D syntheses could not reproduce the 5138\AA\ feature well in 1D. In addition, in most cases the 1D syntheses were broader than the equivalent 3D syntheses. As a result, the 1D fits will tend to underestimate the true value of the heavy isotopes, particularly for \mgii, by up to five percentage points in the most severe case. In effect, \mgi\ is traded for \mgii\ when moving from 1D to 3D. The fraction of \mgiii\ on the other hand, appears to be relatively robust and the true value is essentially recovered for the cases treated here. The magnitude of the differences is increasing with decreasing metallicity. 

We strongly encourage the use of 3D model atmospheres for detailed isotope studies in metal-poor dwarfs, as we find differences that are comparable to, or larger than the typical fitting precision when dealing with real stellar spectra. An increase in the \mgii\ fraction in metal poor stars would result in an increase in both the \mgii/\mgi\ and \mgiii/\mgi\ ratios, which will impact the interpretation of, for instance, the chemical enrichment timescale of the Milky Way halo and the metallicity at which AGB stars become an important contributor to the halo chemistry.

\acknowledgments
HGL acknowledges financial support by the Sonderforschungsbereich SFB\,881 ``The Milky Way System'' (subproject A4) of the German Research Foundation (DFG). AJG acknowledges the support of the FONDATION MERAC, the matching fund granted by the Scientific Council of Observatoire de Paris, and the Collaborative Research Centre SFB 881 (Heidelberg University) of the Deutsche Forschungsgemeinschaft (DFG, German Research Foundation).


\software{\cobold\ \citep{co5bold},  
          Linfor3D \citep{steffen},
          Matplotlib \citep{Hunter:2007},
          Fitprofile \citep{thygesen}
          }



\appendix

\section{Temperature structure of 3D models}
\label{sec:ttau}

\begin{figure*}[htb]
\centering
\includegraphics[width = 0.9\textwidth, trim = 3cm 3cm 3cm 9cm]{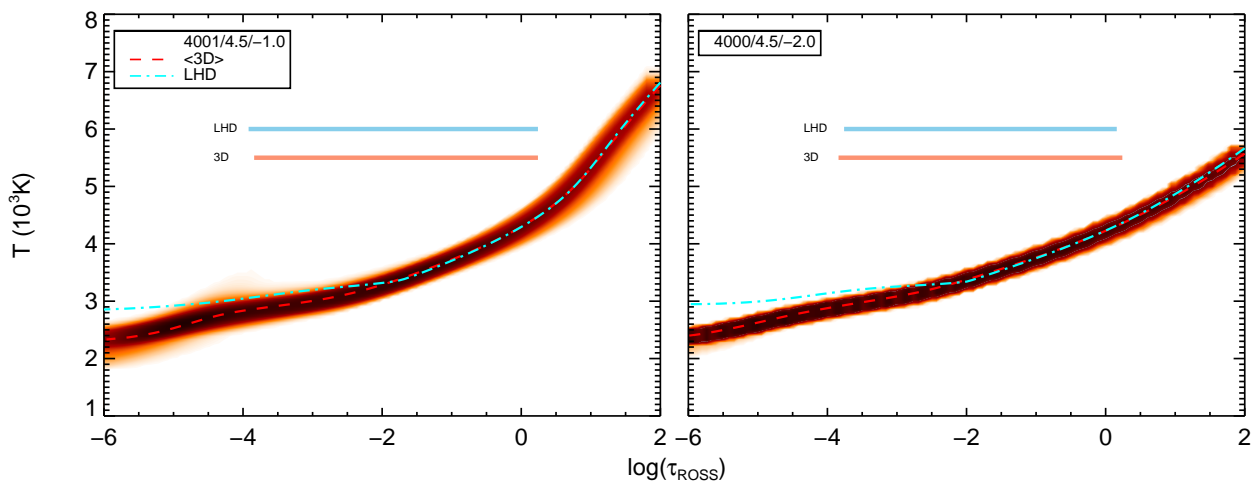}
\caption{The $T(\tau)$ structure of the 4000K \cobold\ models used in this work. Model parameters are given in each panel. The formation regions of MgH in 3D and 1D are indicated as horizontal bars. }%
\label{fig:allTtau_4k}%
\end{figure*}

\begin{figure*}[htb]
\centering
\includegraphics[width = 0.9\textwidth, trim = 3cm 4cm 3cm 3cm]{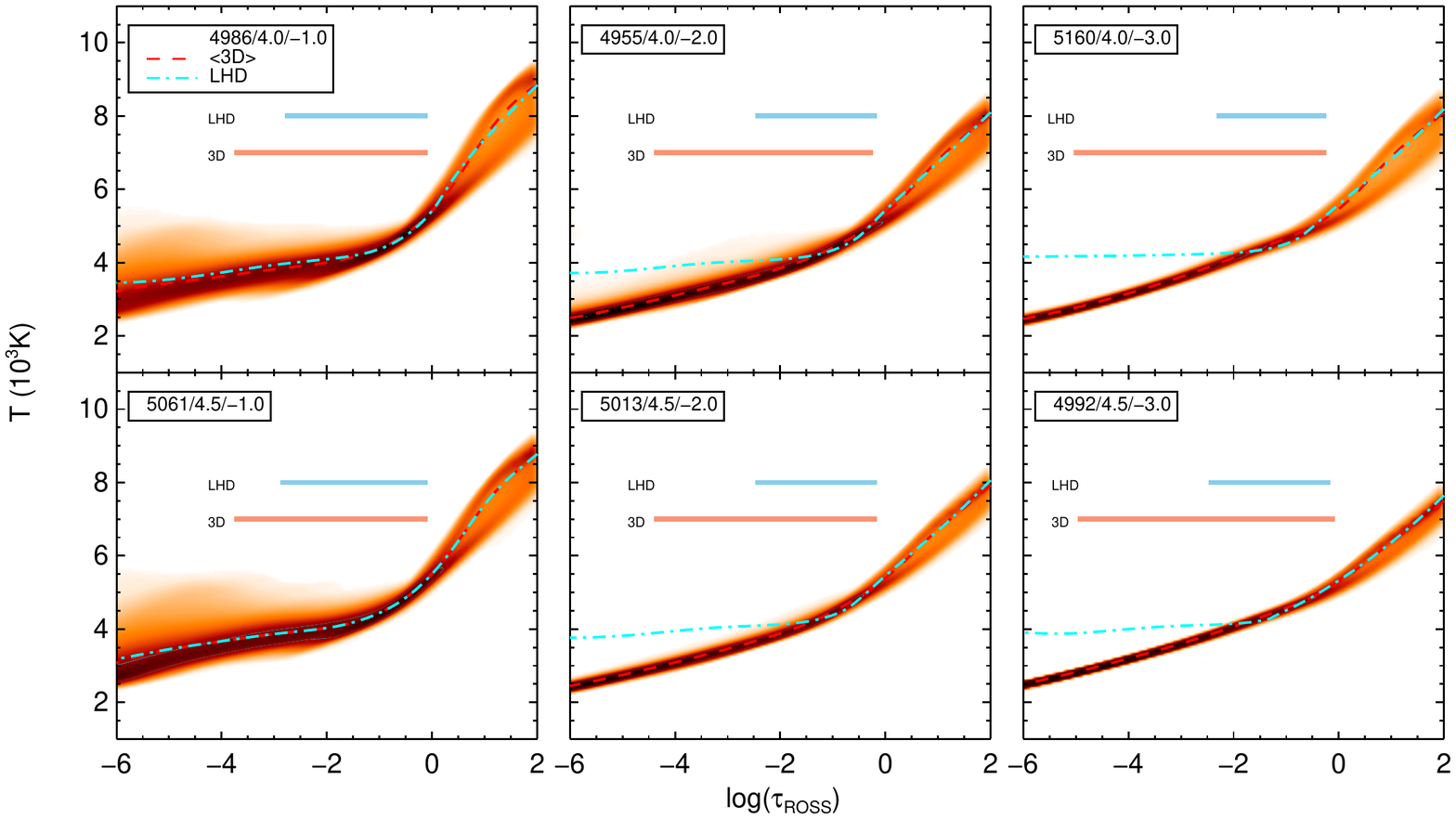}
\caption{The $T(\tau)$ structure of the 5000K \cobold\ models used in this work. Model parameters are given in each panel. The formation regions of MgH in 3D and 1D are indicated as horizontal bars. }%
\label{fig:allTtau_5k}%
\end{figure*}

\section{Number densities of MgH}
\label{sec:nj}

\begin{figure*}[htb]
\centering
\includegraphics[width = 0.9\textwidth, trim = 3cm 4cm 3cm 3cm]{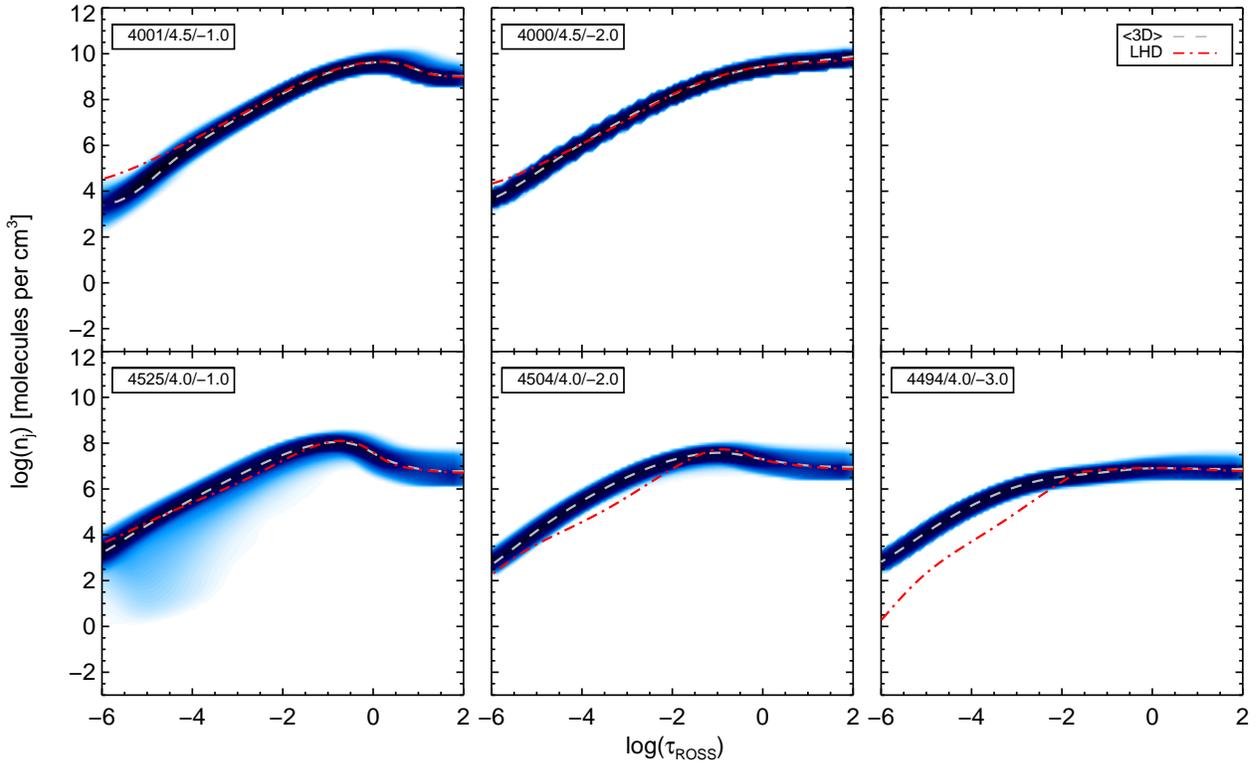}
\caption{The molecular number density, $n_j$, of MgH for the 4000K and 4500K \cobold\ models not presented in Fig.~\ref{fig:mgh-nd}. Model parameters are given in each panel. Darker shaded area indicates a higher sampling of points with this number density.}%
\label{fig:all_nj_4k}%
\end{figure*}

\begin{figure*}[htb]
\centering
\includegraphics[width = 0.9\textwidth, trim = 3cm 4cm 3cm 3cm]{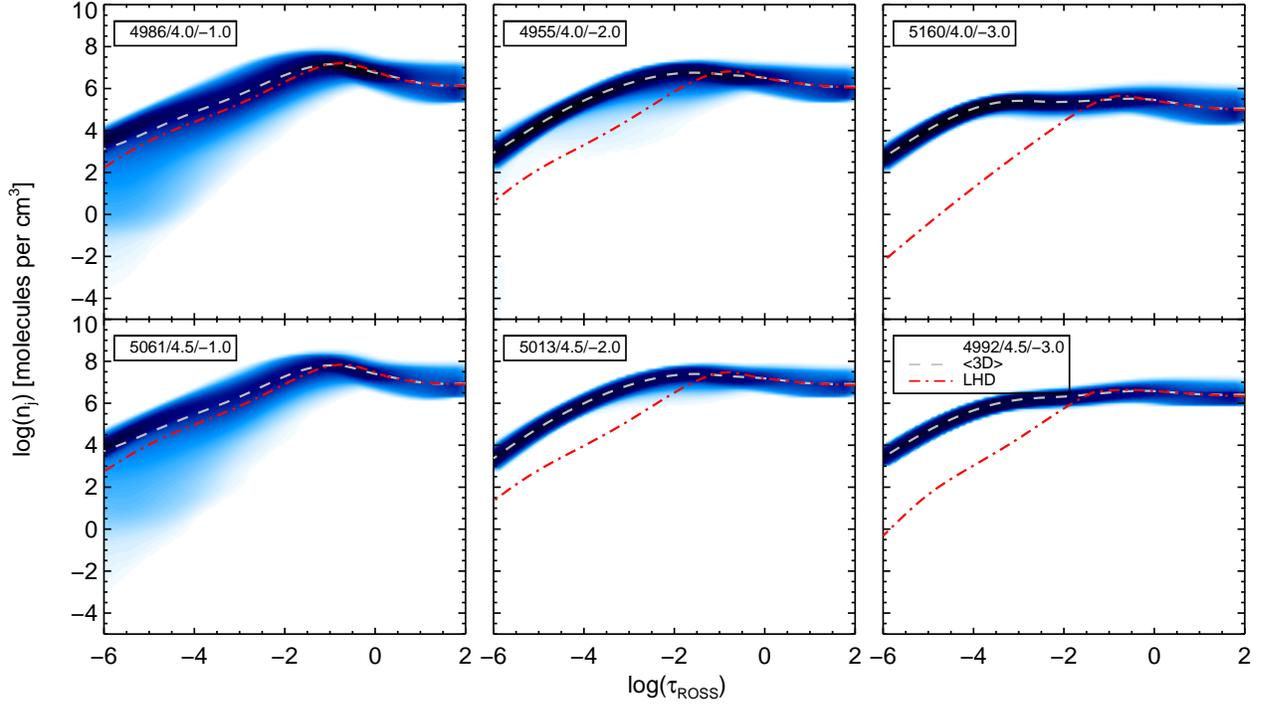}
\caption{The molecular number density, $n_j$, of MgH for the 5000K \cobold\ models used in this work. Model parameters are given in each panel. Darker shaded area indicates a higher sampling of points with this number density.}%
\label{fig:all_nj_5k}%
\end{figure*}





\bibliography{MgH-modeling}




\end{document}